\documentclass[aip, amsmath, amssymb, graphicx, reprint]{revtex4-1}

\usepackage{dcolumn}% Align table columns on decimal point
\usepackage{bm}% bold math
\usepackage[utf8]{inputenc}
\usepackage[T1]{fontenc}
\usepackage{mathptmx}
\usepackage{physics}
\usepackage{dsfont}
\usepackage{siunitx}
\usepackage{standalone}
\usepackage{bm}
\usepackage{soul}
\usepackage{multirow}
 \usepackage{booktabs}

\usepackage{accents}
\newlength{\dhatheight}

\newcommand{\ra}[1]{\renewcommand{\arraystretch}{#1}}

\makeatletter
\def\@email#1#2{%
 \endgroup
 \patchcmd{\titleblock@produce}
  {\frontmatter@RRAPformat}
  {\frontmatter@RRAPformat{\produce@RRAP{*#1\href{mailto:#2}{#2}}}\frontmatter@RRAPformat}
  {}{}
}%
\makeatother

\draft % marks overfull lines with a black rule on the right

\begin{document}

\title{Simulating Non-Markovian Dynamics in Multidimensional Electronic Spectroscopy via Quantum Algorithm}

\author{F. Gallina}
\affiliation{Dipartimento di Scienze Chimiche, Università degli Studi di Padova, via Marzolo 1, Padua 35131, Italy}

\author{M. Bruschi}
\affiliation{Dipartimento di Scienze Chimiche, Università degli Studi di Padova, via Marzolo 1, Padua 35131, Italy}

\author{R. Cacciari}
\affiliation{Dipartimento di Scienze Chimiche, Università degli Studi di Padova, via Marzolo 1, Padua 35131, Italy}

\author{B. Fresch}
\affiliation{Dipartimento di Scienze Chimiche, Università degli Studi di Padova, via Marzolo 1, Padua 35131, Italy}
\affiliation{Padua Quantum Technologies Research Center, Università degli Studi di Padova, via Gradenigo 6/A, Padua 35131, Italy}

\email[email: ]{federico.gallina@unipd.it, barbara.fresch@unipd.it}

\date{\today}

\begin{abstract}
Including the effect of the molecular environment in the numerical modeling of time-resolved electronic spectroscopy remains an important challenge in computational spectroscopy. In this contribution, we present a general approach for the simulation of the optical response of multi-chromophore systems in a structured environment and its implementation as a quantum algorithm. A key step of the procedure is the pseudomode embedding of the system-environment problem resulting in a finite set of quantum states evolving according to a Markovian quantum master equation. This formulation is then solved by a collision model integrated into a quantum algorithm designed to simulate linear and nonlinear response functions. The workflow is validated by simulating spectra for the prototypical excitonic dimer interacting with fast (memoryless) and finite-memory environments. The results demonstrate, on the one hand, the potential of the pseudomode embedding for simulating the dynamical features of nonlinear spectroscopy, including lineshape, spectral diffusion, and relaxations along delay times. On the other hand, the explicit synthesis of quantum circuits provides a fully quantum simulation protocol of nonlinear spectroscopy harnessing the efficient quantum simulation of many-body dynamics promised by the future generation of fault-tolerant quantum computers.
\end{abstract}

\maketitle %\maketitle must follow title, authors, abstract and \pacs

\section{Introduction}
In the last decade, the challenge of disentangling and correctly assigning spectral features to dynamical processes triggered a strong interaction between theoretical models, computational simulations and experimental developments \cite{Ress2023, Sun2024}.
Despite continuous and significant progress, the simulation of the time-resolved spectroscopic response of multi-chromophore systems remains a formidable task at least for two fundamental reasons:
firstly, electronic transitions are often coupled to selected molecular vibrations (vibronic transitions) which produce characteristic patterns in the spectra \cite{Segatta2023}. Because the dimension of the vibronic wavefunction diverges exponentially with the number of relevant vibrational modes, there is an intrinsic scaling issue due to the quantum mechanical nature of the molecular degrees of freedom involved in the system dynamics.
The second major challenge emerges because the spectroscopic response is highly sensitive to the molecular environment, including the multitude of intramolecular vibrations, scaffold and solvent degrees of freedom. The explicit inclusion of this continuum of modes into a numerical protocol is impossible, therefore the theory of open quantum systems is a key theoretical framework in the modeling of the spectroscopic response based on the reduced density matrix of the relevant degrees of freedom.
In this context, perturbative approaches assuming chromophore-environment (Redfield) or chromophore-chromophore (Förster) weak coupling are usually employed to obtain tractable solutions for the complex response of extensive systems and materials \cite{
%Abramavicius2009, Abramavicius2011, 
Valkunas2013}.
However, these assumptions are easily violated in exciton dynamics \cite{Ishizaki2009, Ishizaki2010}, thus requiring numerically more demanding methods to address non-perturbative and non-Markovian parameter regimes \cite{Tanimura1989, Tanimura2020, Makri1995, Bose2020, Chin2013, Lorenzoni2024, Schroter2015}.
Recent efforts on the computational modeling of nonlinear spectroscopy aim to provide resource-efficient simulations of multi-chromophore quantum dynamics \cite{Jansen2021}. Along this line, different approaches have been developed like simulations based on generalized quantum master equations \cite{Sayer2024}, hierarchy of pure states (HOPS) \cite{Chen2022,Chen2022a}, adaptive HOPS \cite{Gera2023} and coarse-grained simulations based on the numerical integration of the Schrödinger equation (NISE) \cite{Zhong2024}, to name a few.

In this paper, we discuss a simulation protocol for the spectroscopic response of a multi-chromophore network, tackling both the fundamental challenges discussed above. We do so in the framework of the emerging technology of quantum computing, as it promises a substantial advantage in the simulation of many-body quantum dynamics \cite{Lloyd1996a, Daley2022}. The potentials of quantum computing for molecular sciences are vast \cite{Weidman2024}, however quantum simulation of spectroscopic responses and parameters has started receiving attention only recently \cite{Sawaya2019, Cai2020, Francis2020, Lee2021, Huang2022, Seetharam2023, OlarteHernandez2024}.
In a previous work \cite{Bruschi2024}, we introduced a quantum algorithm for computing the nonlinear response function based on the quantum simulation of the exciton-vibrational dynamical pathways triggered by the light-matter interaction (also known as double-sided Feynman diagrams, FDs). 
The quantum algorithm is structured as a generalized Hadamard test \cite{Somma2002, Pedernales2014, Wecker2015, Xin2017, Roggero2019, DelRe2022, Chiesa2019, Kale2024}.
The encoding of the exciton-vibrational wavefunction into a qubit register, where the computational space grows exponentially with the number of qubits, allows the efficient simulation of the combined excitonic and vibrational dynamics whose complexity remains polynomial in the number of degrees of freedom defining the molecular system. Here, we elaborate further on the quantum simulation of nonlinear spectroscopies by accounting for the role of the environment with an efficient quantum simulation of the open system dynamics. As a result, we demonstrate quantum circuits to simulate 2D electronic spectra of excitons interacting with a structured molecular environment without assuming weak coupling or Markovian dynamics.

\begin{figure*}
    \centering
    \includegraphics[width=\textwidth]{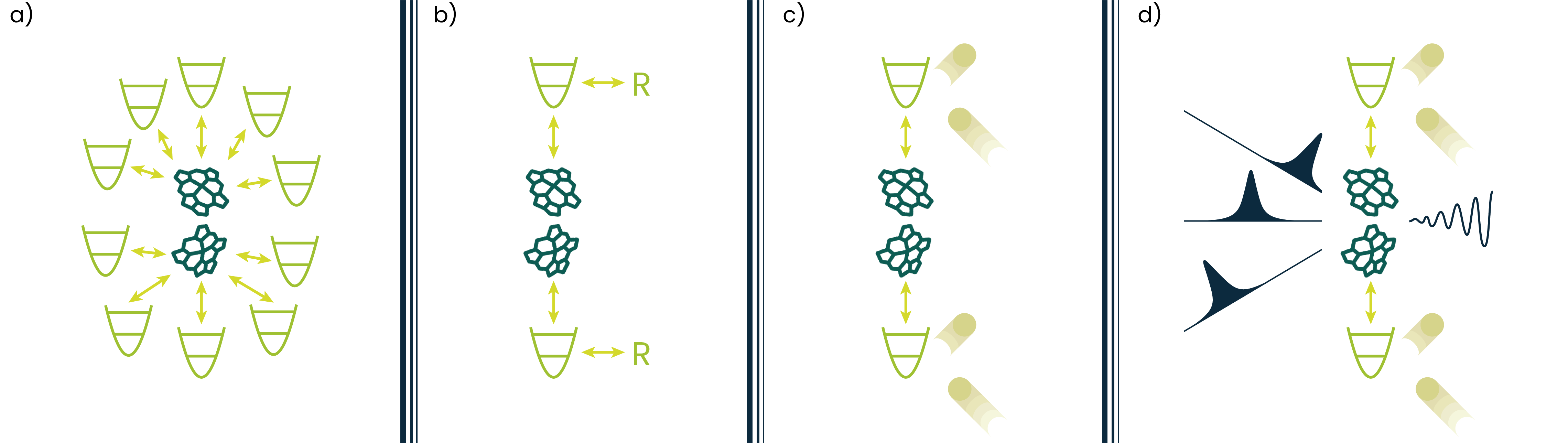}
    \caption{Theoretical methodologies discussed in this manuscript. (a) Physical model of a Frenkel exciton aggregate interacting with a continuum of vibrational degrees of freedom. (b) Markovian embedding of the problem reduces the dimensionality of the environment by substituting it with a finite number of dissipative quantum harmonic oscillators (pseudomodes). (c) The Markovian master equation that describes the system-pseudomode dynamics is mapped to a collision model in which the pseudomodes repeatedly interact with ancillary degrees of freedom. (d) The dynamics is inserted into a quantum algorithm for the simulation of the linear and nonlinear spectroscopic response of the aggregate.}
    \label{fig:Megafigura}
\end{figure*}

The quantum simulation protocol integrates several theoretical methodologies as illustrated in Fig. \ref{fig:Megafigura}.
The starting point is a physical model for the overall system, that is a Frenkel exciton system interacting with a thermal bosonic environment which typically contains a continuum of oscillatory modes (Fig. \ref{fig:Megafigura}a).
Amongst the theoretical frameworks to handle non-Markovian dynamics, the pseudomode model consists of replacing the continuum of environmental degrees of freedom with a few dissipative quantum harmonic oscillators that are chosen to approximate the influence of the original environment on the system dynamics \cite{Imamoglu1994, Imamoglu1994a, Garraway1997, Mazzola2009, Tamascelli2018, Lambert2019, Pleasance2020, Teretenkov2019} (Fig. \ref{fig:Megafigura}b). This equivalent environment is defined by matching the spectral function of the original environment with a discrete set of Lorentzian functions. Formally, each Lorentzian function corresponds to a pseudomode interacting with its zero-temperature Markovian environment \cite{Tamascelli2018}. This procedure effectively transforms a non-Markovian problem into a Markovian one, meaning that the dynamics of the enlarged system, i.e. excitons plus pseudomodes, is described by a master equation in the Lindblad form. We should now point out that the solution of a Lindblad master equation in a quantum computer needs to face the challenge of simulating non-unitary dynamics by means of unitary gates and measurement operations. Here, we propose to solve this problem by setting a collision model implementing the Markovian reservoir via an ancilla qubit (Fig. \ref{fig:Megafigura}c). Collision models were used before for classical simulation of exciton dynamics in biologically inspired open quantum systems \cite{Pedram2022, Chisholm2021} and a collision-based quantum algorithm for Markovian exciton dynamics is presented in Ref. \citenum{Gallina2022}. In a collision model, the relaxation dynamics occurs through repeated interactions between the target system and ancillary two-level systems \cite{Ciccarello2022, Lorenzo2017, Kretschmer2016}, providing an efficient approach that is naturally suited to digital quantum simulation \cite{Kretschmer2016, Garcia-Perez2020, Cattaneo2021, Cattaneo2023}.  Overall, the Markovian embedding obtained with the pseudomodes and the collisional solution of the resulting Lindblad master equation provides a general dynamical model that can be readily translated into a quantum circuit for the simulation of the free dynamics of the exciton system.\\
To obtain the simulation of spectroscopic response, these free-dynamics blocks must be integrated into a quantum circuit describing excitation and de-excitation due to the interaction with the external fields (Fig. \ref{fig:Megafigura}d).
This part of the algorithm, already discussed in Ref. \citenum{Bruschi2024}, allows the components of the nonlinear response function to be recovered by measuring a single ancilla qubit. The flexibility of the algorithm enables the incorporation of the pseudomodes and collision model while keeping the scaling polynomial in the number of chromophores.
 
The rest of the manuscript is organized as follows: in Section \ref{sec:Theory}, we give the theoretical basis underlying the simulation protocol, including an essential formulation of the nonlinear response of a Frenkel exciton model interacting with a structured environment, the definition of the equivalent system in terms of a discrete set of pseudomodes and the collisional implementation of the resulting Markovian dynamics. In Section \ref{sec:Results}, we demonstrate the overall procedure by discussing the results obtained by simulating the linear and nonlinear response of a single chromophore and a molecular dimer undergoing Markovian and non-Markovian dynamics. We focus on a standard model for the environment, such as the overdamped Brownian oscillator, to validate the simulation procedure. On the one hand, we recover the spectral features associated with different spectral densities of the environment. On the other hand, we test the quantum simulation (run in Qiskit Aer) against the classical numerical solution of the same models. In the following Section \ref{sec:Methods}, we discuss the details of the quantum circuits running the simulation and the algorithmic scaling as the number of chromophores increases. We conclude with a brief discussion on the possible source of errors and potential further developments in the quantum simulation of spectroscopy. We provide Appendices containing several technical notes. The code utilized to generate the findings presented in this manuscript is freely accessible in our QSExTra \cite{QSExTra} module for Python.

%%%%%%%%%%%%%%%%%%%%%%%%%%%%%%%%%%%%%%%%%%%%%%%%%%%%%%%%%%%%%%%%%%%%%%%%
%%%%%%%%%%%%%%%%%%%%%%%%%%%%%%%%%%%%%%%%%%%%%%%%%%%%%%%%%%%%%%%%%%%%%%%%
\section{\label{sec:Theory}Theory}
%%%%%%%%%%%%%%%%%%%%%%%%%%%%%%%%%%%%%%%%%%%%%%%%%%%%%%%%%%%%%%%%%%%%%%%%
\subsection{A multi-chromophore aggregate in a molecular environment}
The border between the system of interest and the environment is traced by partitioning the overall Hamiltonian. Here, the system is intended as the excitonic degrees of freedom of a multi-chromophore aggregate naturally surrounded by vibration, rotation, libration, and solvent degrees of freedom which form the molecular environment.
We define the global Hamiltonian as
\begin{equation} \label{eq:H}
    \hat{H}^\text{0} = \hat{H}^\text{S} + \hat{H}^\text{E} + \hat{H}^\text{SE}
\end{equation}\
which is composed of the system Hamiltonian $\hat{H}^\text{S}$, the environment Hamiltonian $\hat{H}^\text{E}$ and their interaction $\hat{H}^\text{SE}$.

The system is described by a Frenkel Hamiltonian
\begin{equation} \label{eq:H_S}
    \hat{H}^\text{S} = - \sum_{i=1}^N \frac{\epsilon_i}{2} \hat{\sigma}_i^z + \sum_{i=1}^{N-1} \sum_{j>i} J_{ij} \left( \hat{\sigma}_i^+\hat{\sigma}_j^- + \hat{\sigma}_i^- \hat{\sigma}_j^+ \right),
\end{equation}
where the $N$ chromophores are characterized by an electronic energy gap $\epsilon_i$ and are coupled pairwise by Coulomb interaction terms $J_{ij}$.
Each chromophore is considered as a two-level system with an electronic ground state $\ket{g_i}$ and a first excited state $\ket{e_i}$.
In eq. \ref{eq:H_S}, the system operators are written in terms of the Pauli-z operator $\hat{\sigma}_i^z = \ketbra{g_i}{g_i} - \ketbra{e_i}{e_i}$ and ladder operators $\hat{\sigma}_i^+ = \ketbra{e_i}{g_i}$ and $\hat{\sigma}_i^- = \ketbra{g_i}{e_i}$.

The environment is modeled by a collection of independent quantum harmonic oscillators with Hamiltonian
\begin{equation} \label{eq:H_E}
    \hat{H}^\text{E} = \sum_{i=1}^N \sum_{k} \omega_{ik} \hat{b}_{ik}^\dagger \hat{b}_{ik}
\end{equation}
where $\omega_{ik}$ is the natural frequency, and $\hat{b}_{ik}^\dagger$ and $\hat{b}_{ik}$ are the bosonic creation and annihilation operators of the $k$-th mode associated with chromophore $i$.
The system-environment interaction is regulated by
\begin{equation} \label{eq:H_SE}
    \hat{H}^\text{SE} = \sum_{i=1}^N \ketbra{e_i}{e_i} \hat{X}_i
\end{equation}
with
\begin{equation}
    \hat{X}_i = \sum_{k} c_{ik} \left( \hat{b}_{ik}^\dagger + \hat{b}_{ik} \right)
\end{equation}
where $c_{ik}$ the coupling strenght between a mode $k$ and a chromophore $i$.

In principle, the system-environment dynamics can be described using the Schrödinger equation for the global statevector, or the equivalent formulation in terms of the density matrix $\rho(t)$ evolving according to the Liouville-von Neumann equation (we set $\hbar = 1$)
\begin{equation} \label{eq:sys-bigenv_vonNeumann}
    \frac{d \rho(t)}{dt} = - i [\hat{H}^0, \rho(t)],
\end{equation}
whose solution defines the unitary propagator
\begin{equation} \label{eq:sys-bigenv_unitary_dynamics}
    \rho(t + t') = \mathcal{U}_{t'} [\rho(t)] = e^{-i\hat{H}^0t'} \rho(t) e^{i\hat{H}^0t'}.
\end{equation}
%
%%%%%%%%%%%%%%%%%%%%%%%%%%%%%%%%%%%%%%%%%%%%%%%%%%%%%%%%%%%%%%%%%%%%%%%%
\subsection{Spectroscopic response of the system}
According to response theory, a system interacting $M$ times with an electric field $E(t)$ emits a signal that is proportional to the $M$-th order component of the perturbative expansion of the polarization \cite{Mukamel1995}
\begin{equation} \label{eq_polarization}
\begin{split}
    P^{(M)}(t) =& \int_0^\infty dt_M \dots \int_0^\infty dt_1 R^{(M)}\left( t_M, \dots, t_1 \right) \times\\
    & \times E(t-t_M) \dots E(t-t_M \dots -t_1)
\end{split}
\end{equation}
where $t_m = \tau_{m+1} - \tau_{m}$ is the delay time between two consecutive interactions, and $t_M$ is the delay time between the last interaction and the signal emission.
Since $E(t)$ is known, one is interested in calculating the $M$-th order system response function $R^{(M)}$ to interpret the signal \cite{Mukamel1995}.
This has the form
\begin{multline} \label{eq:response_function}
    R^{(M)}\left( t_M, \dots, t_1 \right) =
    \left( i \right)^M
    \Tr_\text{S}\Big\{
    \hat{\mu}
    \mathcal{U}_{t_M} \circ
    \mathcal{M} \circ
    \dots \\
    \dots
    \circ
    \mathcal{U}_{t_1} \circ
    \mathcal{M} \left[
    \rho^{(0)}(0)
    \right]
    \Big\}
\end{multline}
where the notation $\mathcal{A} \circ \mathcal{B} \left[\cdot \right] = \mathcal{A} \left[ \mathcal{B} \left[\cdot \right] \right]$ indicates the composition of dynamical maps and $\rho^{(0)}(0) = \rho(0)$ is the system-environment state prior to the external perturbation, which is supposed to be the product state between the system ground state and a thermal state of the environment.
Notice that we are assuming that the system-environment dynamics along the delay times $\mathcal{U}_{t_m}$ is not influenced by the presence of the electric field and that the composition property $\mathcal{U}_{t+t'} = \mathcal{U}_{t'} \circ \mathcal{U}_{t}$ holds (which is always true for unitary and Markovian dynamics).
The superoperator $\mathcal{M}$ represents the effect of the light-matter interaction, corresponding to the commutator $\mathcal{M}[\rho^{(m)}(t_m)] = \left[ \hat{\mu}, \rho^{(m)}(t_m) \right]$ with the transition dipole moment operator $\hat{\mu}$. Under the Frank-Condon approximation, we can decompose the transition dipole moment as $\hat{\mu} = \hat{\mu}^+ + \hat{\mu}^-$, where the components
\begin{equation} \label{eq:dipole_operators}
    \hat{\mu}^\pm = \sum_{i=1}^N \mu_i \hat{\sigma}_i^\pm
\end{equation}
regulate the excitation (+) and de-excitation (-) of the aggregate and $\{\mu_i\}$ are the strengths of the transition dipole moments of the chromophores.
The application of $\mathcal{M}$ updates the system state to the subsequent order in the perturbative expansion $\rho^{(m+1)}(t_m) = \mathcal{M}[\rho^{(m)}(t_m)]$.

\begin{figure*}
    \centering
    \includegraphics[]{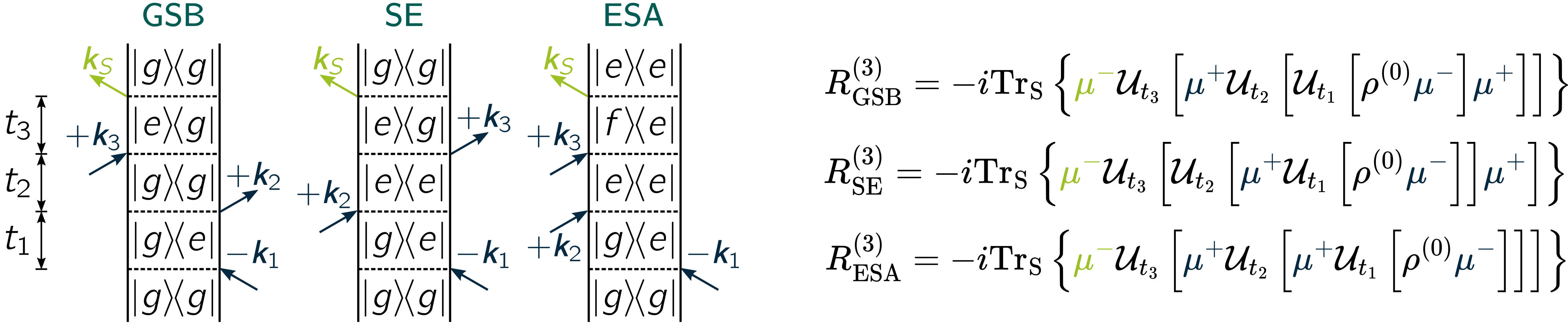}
    \caption{Double-sided Feynman diagrams and response functions of the third-order ground-state bleaching, stimulated emission, and excited-state absorption of a single chromophore. In a double-sided Feynman diagram, time flows from bottom to top. The two vertical lines represent the ket (left) and bra (right) sides of the system density matrix. The direction of the arrows indicates the excitation (incoming) or de-excitation (outgoing) of the system resulting from the light-matter interaction. The (green) arrow on the top of the diagram represents the signal emission.}
    \label{fig:FDs_example}
\end{figure*}

By expanding the commutators, one obtains the response function as the sum over $2^M$ terms, each describing a possible dynamical pathway followed by the system during the coherent excitation
\begin{equation}
    R^{(M)}\left( t_M, \dots, t_1 \right) =
    \sum_{\alpha = 1}^{2^M}
    R^{(M)}_\alpha\left( t_M, \dots, t_1 \right).
\end{equation}
Under the rotating wave approximation, only one of the two components of the dipole moment operator gives non-negligible effects, and we have
\begin{multline} \label{eq:response_function_alpha}
    R^{(M)}_\alpha\left( t_M, \dots, t_1 \right) =
    s_\alpha
    \left( i \right)^M
    \Tr_\text{S}\Big\{
    \hat{\mu}^-
    \mathcal{U}_{t_M} \circ
    \mathcal{V}_M^\alpha \circ
    \dots \\
    \dots
    \circ
    \mathcal{U}_{t_1} \circ
    \mathcal{V}_1^\alpha \left[
    \rho^{(0)}(0)
    \right]
    \Big\}
\end{multline}
where $s_\alpha$ is a sign arising from the commutators and 
\begin{equation} \label{eq:light-matter_map}
    \mathcal{V}_m^\alpha \left[\rho^{(m-1)}_\alpha(\tau_m) \right] =
\begin{cases}
    \hat{\mu}^\pm \ \rho^{(m-1)}_\alpha(\tau_m)
    %\ \hat{\mathbb{I}}^\text{S}
    ,\\
    \\
    %\hat{\mathbb{I}}^\text{S} \ 
    \rho^{(m-1)}_\alpha(\tau_m) \ \hat{\mu}^\pm
\end{cases}
    = \rho^{(m)}_\alpha(\tau_m)
\end{equation}
which indicates the update of the density matrix due to the interaction with the rotating or counter-rotating component of the electric field.
The sequence of (bra- and ket-side) oriented applications of the dipole moments can be visualized using a FD, which is the graphical representation of the dynamical pathway.
In Fig. \ref{fig:FDs_example}, we show the FDs of the third-order Ground-State Bleaching (GSB), Stimulated Emission (SE), and Excited-State Absorption (ESA) of a single chromophore with the related response functions.

As already discussed in Ref.\citenum{Bruschi2024}, the structure of FDs suggests a quantum algorithm for simulating the system response function based on alternating the oriented application of dipole operators and field-free time evolutions.
Therefore, one can simulate the response function based on the system-environment time propagation, according to eq. \ref{eq:sys-bigenv_unitary_dynamics}. 
However, a typical assumption is that the bosonic environment is composed of a continuum of modes.
Although this is the case where the quantum advantage is more explicit, the quantum implementation can become demanding in terms of qubit embedding and depth of the circuit encoding the system-environment dynamics.
To address the curse of dimensionality without losing the generality of the description, we next introduce a model to reshape the environment into a minimal set of effective degrees of freedom.

%%%%%%%%%%%%%%%%%%%%%%%%%%%%%%%%%%%%%%%%%%%%%%%%%%%%%%%%%%%%%%%%%%%%%%%%
\subsection{Markovian embedding with pseudomodes}
If we assume that the initial state of eq. \ref{eq:sys-bigenv_vonNeumann} is a product state between the system and the environment and that the environment is initially found in a Gaussian state (such as a thermal state), then Wick's theorem ensures that the effect of the environment on the system dynamics is fully characterized by its two-time correlation function
$
C_i(t) = \langle \hat{X}_i(t) \hat{X}_i(0) \rangle
$,
where $\hat{X}_i(t) = e^{i\hat{H}^\text{E}t} \hat{X}_i e^{-i\hat{H}^\text{E}t}$.
As a consequence, any Gaussian environment with the same correlation function affects the open system in the same way.

\begin{figure}
    \centering
    \includegraphics[width=0.48\textwidth]{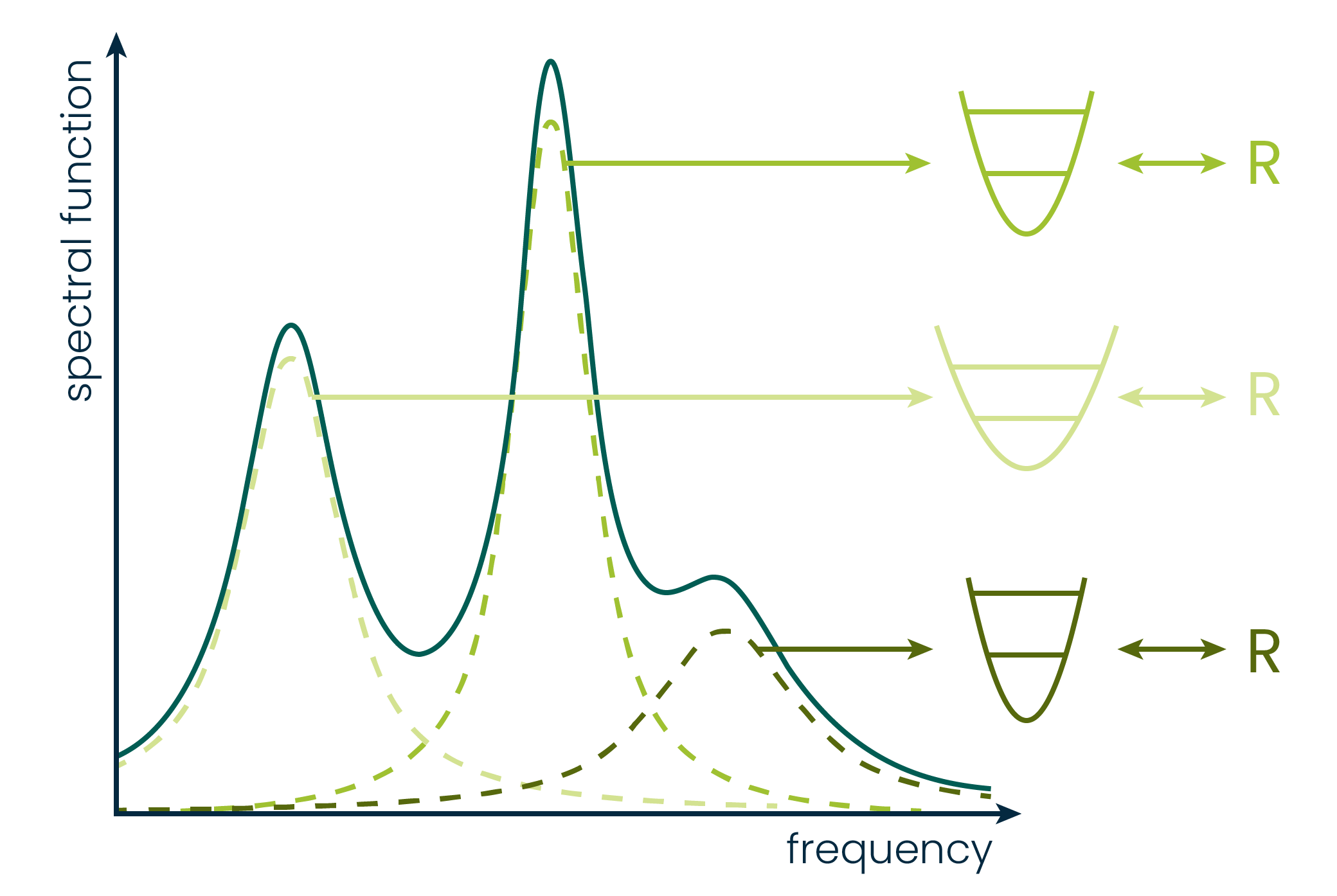}
    \caption{Fitting of a spectral function with Lorentzians. In the pseudomode model, each Lorentzian function represents a quantum harmonic oscillator coupled to the system and a Markovian reservoir at zero temperature.}
    \label{fig:lorentzian}
\end{figure}

For the sake of simplicity, in the following, we will consider the environment of each chromophore to have the same correlation function, i.e., $C_i(t) = C(t)$.
Furthermore, we assume that the spectral function $C(\omega)$ of the environment, corresponding to the Fourier transform of $C(t)$, can be fitted with a sum of $W$ Lorentzians \cite{Leppakangas2023, Mascherpa2020}
\begin{equation} \label{eq:spectral_function}
    C(\omega) = \int_{-\infty}^\infty dt\ e^{i \omega t} C(t)
    \approx \sum_{k=1}^W \Gamma_k \frac{\Omega_k^2}{\left( \omega - \omega_k^\text{0} \right) ^2 + \Omega_k^2 },
\end{equation}
with a positive amplitude $\Gamma_k$, width $\Omega_k$ and central frequency $\omega_k^\text{0}$ (Fig. \ref{fig:lorentzian}).
Now, we can exploit the bath Gaussianity together with the Lorentzian decomposition to embed the continuous environment into a discrete number of effective modes called pseudomodes \cite{Tamascelli2018}.

Pseudomodes are quantum harmonic oscillators coupled to the system and a zero-temperature Markovian reservoir.
The initial state of the pseudomodes is the factorized state $\rho_\text{P}(0) = \bigotimes_{i=1}^N \bigotimes_{k=1}^W \ketbra{0_{ik}}{0_{ik}}$.
We define the new system-pseudomode Hamiltonian as
\begin{equation} \label{eq:H_tot}
    \hat{H} = \hat{H}^\text{S} + \hat{H}^\text{P} + \hat{H}^\text{EP},
\end{equation}
where the pseudomode Hamiltonian is
\begin{equation} \label{eq:H_HO}
    \hat{H}^\text{P} = \sum_{i=1}^N \sum_{k=1}^W \omega_{k} \hat{a}_{ik}^\dagger \hat{a}_{ik}
\end{equation}
with $\hat{a}_{ik}^\dagger$ and $\hat{a}_{ik}$ the creation and annihilation operators of a pseudomode and $\omega_{k}$ its frequency.
The system-pseudomode interaction Hamiltonian is regulated by
\begin{equation} \label{eq:H_int}
    \hat{H}^\text{EP} = \sum_{i=1}^N \ketbra{e_i}{e_i} \sum_{k=1}^W g_{k} \left( \hat{a}_{ik}^\dagger + \hat{a}_{ik} \right).
\end{equation}
with coupling strenght $g_{k} = \sqrt{\Gamma_k \Omega_k / 2}$.

By identifying $\omega_{k} = \omega_k^\text{0}$, each pseudomode contributes to the spectral function with one peak centered at a desired frequency.
The Lorentzian broadening is regulated by the interaction with the Markovian reservoir.
In this context, the system-pseudomodes density matrix $\rho (t)$ evolves under the master equation
\begin{equation} \label{eq:master_equation}
    \frac{d \rho (t)}{dt} =
    -i \left[ \hat{H}, \rho (t) \right] +
    \mathcal{D} \left[ \rho (t) \right]
\end{equation}
with Lindblad-like dissipators
\begin{equation} \label{eq:dissipator}
    \mathcal{D} \left[ \rho (t) \right] =
    \sum_{i=1}^N
    \sum_{k=1}^W 2\Omega_k
    \left(
    \hat{a}_{ik} \rho (t) \hat{a}_{ik}^\dagger -
    \frac{1}{2} \left[ \hat{a}_{ik}^\dagger \hat{a}_{ik}, \rho (t) \right]_+
    \right),
\end{equation}
where $[A,B]_+$ is the anticommutator.
Inside the dissipator (eq. \ref{eq:dissipator}), $2 \Omega_k$ represents the relaxation rate constants of a pseudomode $k$.
In the limit of $\Omega_{\overline{k}} \to \infty$, i.e., when the relaxation time is short compared to the system timescale, the corresponding pseudomode $\overline{k}$ is always in the vacuum state during the system dynamics. In this case, pseudomode contribution is a flat spectral function. Therefore, the degrees of freedom associated with the pseudomode can be omitted from the Hamiltonian, as its influence is easily captured by the dissipator \cite{Gallina2024}
\begin{equation} \label{eq:dissipator_memoryless}
\begin{split}
    \mathcal{D}_{\overline{k}} \left[ \rho(t) \right] &=
    \sum_{i=1}^N \Gamma_{\overline{k}} \left( \ketbra{e_i}{e_i} \rho(t) \ketbra{e_i}{e_i} - \frac{1}{2} \left[ \ketbra{e_i}{e_i}, \rho(t) \right]_+ \right)\\
    &=\sum_{i=1}^N \frac{\Gamma_{\overline{k}}}{4} \big( \hat{\sigma}_i^z \rho(t) \hat{\sigma}_i^z - \rho(t) \big).    
\end{split}
\end{equation}
On the other hand, when the relaxation time is comparable to or longer than the system timescale, the relaxation of the pseudomode is sufficiently slow to retain information about past excitonic states, thereby influencing the dynamics of the open system.
In this scenario, the pseudomode acts effectively as a memory kernel.
%%%%%%%%%%%%%%%%%%%%%%%%%%%%%%%%%%%%%%%%%%%%%%%%%%%%%%%%%%%%%%%%%%%%%%%%
\subsection{A collision model for the Markovian reservoir}
Since quantum computers rely on unitary operations and measurements, we need a framework for performing the non-unitary dynamics of the open quantum system in terms of quantum circuits.
We do so by setting a collision model, which can efficiently reproduce our target Lindblad master equation.

The evolution of the system-pseudomode state $\rho(t)$ is discretized in small time intervals of length $\delta t$.
The update of the system state from time $t$ to $t+\delta t$ occurs in two steps \cite{Lorenzo2017}: Firstly, the system plus pseudomodes evolves under the Hamiltonian propagator
\begin{equation}\label{eq:Hamiltonian propagator}
    \hat{U}(\delta t) = e^{- i \hat{H} \delta t},
\end{equation}
then the pseudomodes interact (``collide") with ancillary two-level systems.
Each collision involves a pseudomode and an ancilla in the initial ground state $\ket{0_\text{C}}$.
The collision operator is
\begin{equation}\label{eq:collision propagator}
    \hat{U}^\text{C}(\delta t) = \prod_{i=1}^N \prod_{k=1}^W \hat{U}_{ik}^\text{C}(\delta t) = \prod_{i=1}^N \prod_{k=1}^W e^{- i \hat{H}_{ik}^\text{C} \delta t},
\end{equation}
with collision Hamiltonian
\begin{equation}
    \hat{H}_{ik}^\text{C} \left( \delta t \right) =
    \sqrt{\frac{2 \Omega_k}{\delta t}}
    \left(
    \hat{a}_{ik}^\dagger \hat{\sigma}_\text{C}^- +
    \hat{a}_{ik} \hat{\sigma}_\text{C}^+
    \right),
\end{equation}
where $\hat{\sigma}_\text{C}^\pm$ are the ladder operators of one ancilla.
After the pseudomode-ancilla interactions, the ancillary degrees of freedom can be traced out and eventually replaced with a new set of ancillae for the subsequent evolution of a time step.
The role of the ancillae is thus that of a memoryless (Markovian) environment.
In Section \ref{sec:Methods}, we discuss different implementations of the ancillary degrees of freedom.

Overall, the state-updating equation can be written as
\begin{equation}
    \rho \left( t + \delta t \right) =
    \mathcal{U}_{\delta t} \left[ \rho(t) \right]
\end{equation}
with
\begin{equation}\label{eq:collision time map}
    \mathcal{U}_{\delta t} \left[ \rho(t) \right] =
    \Tr_\text{C}
    \left\{
    \hat{U}^\text{C} \left( \delta t \right)
    \hat{U} \left( \delta t \right)
    %\left( 
    \rho (t) \otimes \rho_\text{C}^0
    %\right)
    \hat{U}^\dagger \left( \delta t \right)
    \hat{U}^{\text{C} \dagger} \left( \delta t \right)
    \right\}.
\end{equation}

In Appendix \ref{app:collision_to_master_equation}, we show that the collision model introduced above reproduces the target master equation (eq. \ref{eq:master_equation}) in the limit of small $\delta t$. Therefore the time evolution over $t_m$ can be performed as
\begin{equation}
\begin{split}
    \mathcal{U}_{t_m} [\rho^{(m)}_\alpha (\tau_m)] &= \lim_{\delta t \to 0} \mathcal{U}_{\delta t}^{t_m/\delta t} \left[ \rho^{(m)}_\alpha (\tau_m) \right] \\
    &\approx \underbrace{\mathcal{U}_{\Delta t} \circ \dots \circ \mathcal{U}_{\Delta t}}_{t_m/\Delta t} \left[ \rho^{(m)}_\alpha (\tau_m) \right].
\end{split}
\end{equation}
where $\Delta t$ is a finite approximation of $\delta t$ such that $t_m/\Delta t \in \mathbb{N}$.

The same procedure can also be applied to the case of the dissipator in eq. \ref{eq:dissipator_memoryless}.
In this case, the collision occurs directly between the chromophore and the ancilla under the Hamiltonian \cite{Gallina2022}
\begin{equation}
    \hat{H}^\text{C}_{i\overline{k}} \left( \delta t \right) = \sqrt{\frac{\Gamma_{\overline{k}}}{4 \delta t}}
    \hat{\sigma}_{i}^z\hat{\sigma}_\text{C}^x.
\end{equation}

Because of its versatility, a collision scheme can be derived for other dissipation processes, such as exciton recombination and trapping, often included in multi-chromophore models \cite{Mohseni2008}.
The collision Hamiltonians for these processes are reported in Appendix \ref{app:relax_and_trap} for completeness, but will not be considered in the following.

The new dynamical map $\mathcal{U}_{t_m} [\rho^{(m)}_\alpha (\tau_m)]$ obtained from the collision model can be easily inserted in eq. \ref{eq:response_function_alpha} and used for calculating the system response function via the quantum algorithm introduced in Section \ref{sec:Methods}.
%%%%%%%%%%%%%%%%%%%%%%%%%%%%%%%%%%%%%%%%%%%%%%%%%%%%%%%%%%%%%%%%%%%%%%%%
%%%%%%%%%%%%%%%%%%%%%%%%%%%%%%%%%%%%%%%%%%%%%%%%%%%%%%%%%%%%%%%%%%%%%%%%
\section{\label{sec:Results}Results}
We now discuss the simulation of the absorption spectrum of an excitonic dimer and the third-order rephasing spectra of both a single chromophore and the dimer.
The results are obtained by simulating the response function with our quantum algorithm and taking its Fourier transform to produce the spectra.
The quantum circuits have been emulated on a classical high-performance computer (HPC) using Qiskit Aer \cite{Qiskit}, as the circuit depth required for the simulations exceeds the noise threshold of today's quantum computer. The results are benchmarked against a standard numerical calculation of the response (eq. \ref{eq:response_function_alpha}) based on the solution of eq. \ref{eq:master_equation} for propagating the dynamics along the delay times.

The parameters of the systems Hamiltonian, the environment, and technical details of the simulation are reported in Tab. \ref{tab:data}.
The excitation energies of the dimer have been chosen to match the optical bands of the B800 and B850 rings of the LH2 complex \cite{Bruschi2024, Damtie2017}.
We simulate both a memoryless and a finite-memory environment in order to compare their signatures in linear and nonlinear optical spectra. The memoryless environment is characterized by a flat spectral function completely characterized by the value of $\Gamma$. The finite-memory environment has a Lorentzian spectral function centered at zero frequency, corresponding to the well-known overdamped Brownian oscillator model at high temperature \cite{Gallina2024, Bondarenko2020}.

\begin{table*}\centering
\ra{1.3}
\begin{tabular}{@{}p{0.2\textwidth}p{0.15\textwidth}p{0.15\textwidth}p{0.15\textwidth}p{0.15\textwidth}@{}}
\toprule
\textbf{System}\\
Single Chromophore & $\epsilon = 1.55$ eV & $-$ & $-$ & $\mu = 1$ a.u.\\
Excitonic Dimer & $\epsilon_1 = 1.55$ eV & $\epsilon_1 = 1.46$ eV & $J_{12} = -0.01$ eV & $\mu_1 = \mu_2 = 1$ a.u.\\
\midrule
\textbf{Environment}\\
Memoryless & $\Gamma = 0.05908$ eV & $(\Omega \to \infty)$ & $-$\\
Finite-memory & $\Gamma = 0.05908$ eV & $\Omega = 0.1$ eV & $\omega^\text{0} = 0$\\
\midrule
\textbf{Simulation details}\\
Propagation time step & $\Delta t = 0.1$ fs\\
Rotating frame & $\omega_{RF} = 1.505$ eV\\
\bottomrule
\end{tabular}
\caption{Parameters used in the simulations of linear and nonlinear spectra.}
\label{tab:data}
\end{table*}

%%%%%%%%%%%%%%%%%%%%%%%%%%%%%%%%%%%%%%%%%%%%%%%%%%%%%%%%%%%%%%%%%%%%%%%%
\subsection{Absorption spectra}
The response function for the linear absorption $R^{(1)}(t_1)$ is composed of two terms $R^{(1)}_1(t_1) = i \Tr\left\{ \hat{\mu}^- \mathcal{U}_{t_1} [\hat{\mu}^+ \rho^{(0)}(0)] \right\}$ and $R^{(1)}_2(t_1) = R^{(1)*}_1(t_1)$.
Since these terms are complex conjugates, it is sufficient to calculate only one of them to obtain the absorption spectra.
In particular, we compute $R^{(1)}_1(t_1)$.

To collect the temporal response function, we scan over the delay time $t_1$ up to $200\ \si{fs}$.
In principle, the time interval $\Delta t$ of the propagation can also be used for scanning the delay time.
However, we use a longer time interval of $10\ \si{fs}$ to reduce the computational burden.
As a side effect, we produce an undersampled signal which can result in shifted spectra frequencies.
In analogy with experimental practices, this can be rectified by applying a rotating frame \cite{Schlau-Cohen2011, Bolzonello2017}, which enables the spectra to be relocated to the appropriate frequency range.

\begin{figure}
    \centering
    \includegraphics[width=0.48\textwidth]{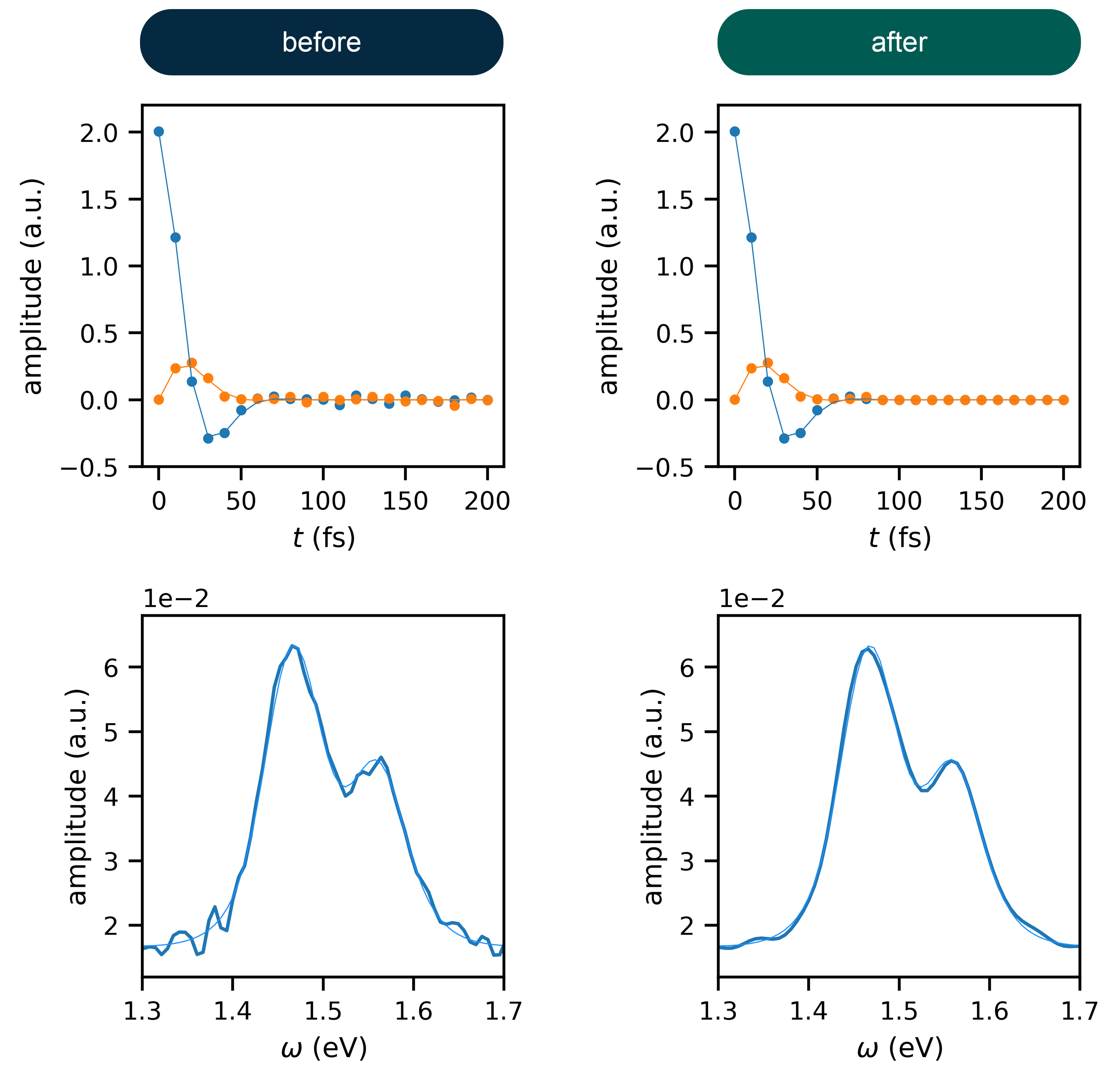}
    \caption{Smoothing procedure for the absorption. The response function $R^{(1)}_1(t_1)$ (in the rotating frame), calculated with the quantum algorithm, displays symptoms of finite-sampling noise, which is particularly evident in the signal tail. This results in the emergence of random frequency contributions within the spectrum. In order to smooth the spectrum, the response is cut where the signal/noise ratio is unfavorable and zero-padded.}
    \label{fig:sampling}
\end{figure}

For each time point, we measure the related circuits with $2 \times 10^4$ sampling shots.
As can be seen from Fig. \ref{fig:sampling}a, this introduces a finite-sampling noise that is particularly evident where the time trace of the response is close to zero.
Notice that this phenomenology occurs because of the loss of correlation due to the system-environment interaction, and was not present in our previous work \cite{Bruschi2024} where the decay of the signal was introduced in the post-processing with an ad hoc window function.
To mitigate the random frequency contributions in the spectrum (Fig. \ref{fig:sampling}b) due to the sampling noise, the signal tail can be cut from the point where the signal-to-noise ratio is lower than one, and substituted with zero-padding (Fig. \ref{fig:sampling}c).
Eventually, the signal is Fourier transformed to obtain the smoothed spectra Fig. \ref{fig:sampling}d.

\begin{figure}
    \centering
    \includegraphics[width=0.48\textwidth]{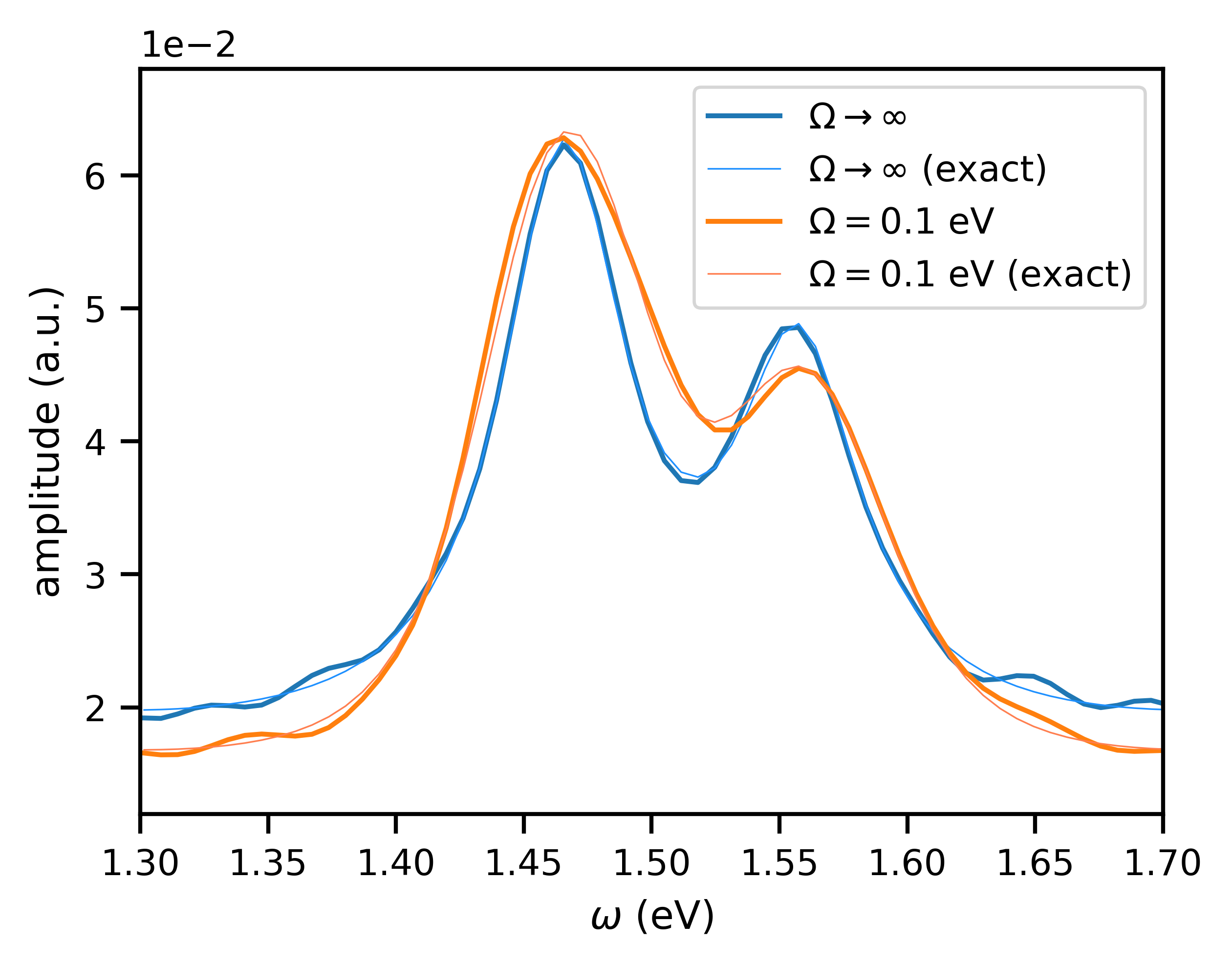}
    \caption{Absorption spectra of the excitonic dimer system in contact with the memoryless (blue) and finite-memory (orange) environments. The results of the quantum algorithm (thick lines) are compared with standard numerical calculations (thin lines).}
    \label{fig:abs}
\end{figure}

In Fig. \ref{fig:abs}, we report the absorption spectra for the excitonic dimer interacting with a memoryless (blue lines) and finite memory environment (orange lines).
The effect of the finite relaxation rate of the pseudomode ($\Omega$) is clearly visible in the different broadenings of the absorption bands that go from the typical Lorentzian to a Kubo lineshape \cite{Kubo1969}.

%
%%%%%%%%%%%%%%%%%%%%%%%%%%%%%%%%%%%%%%%%%%%%%%%%%%%%%%%%%%%%%%%%%%%%%%%%
\subsection{\label{subsec:Nonlinear spectra}Nonlinear spectra}
The third-order rephasing spectrum is constituted by the signal components emitted along the direction $\overrightarrow{k}_\text{S} = - \overrightarrow{k}_1 + \overrightarrow{k}_2 + \overrightarrow{k}_3$.
These are the Ground-State Bleaching (GSB), Stimulated Emission (SE), and Excited-State Absorption (ESA) whose FDs and response functions are shown in Fig. \ref{fig:FDs_example}.

For simulating third-order responses, we scan over three delay times: the coherence times $t_1$ and $t_3$, which are Fourier transformed in the spectra, and the waiting time $t_2$ where we see the relaxation dynamics and energy transfer.
Coherence times $t_1$ and $t_3$ are collected up to $110\ \si{fs}$ with a time intervals of $10\ \si{fs}$.
The undersampled signal recorded is then post-processed as discussed for the absorption.

\begin{figure*}
    \centering
    \includegraphics[width=0.8\textwidth]{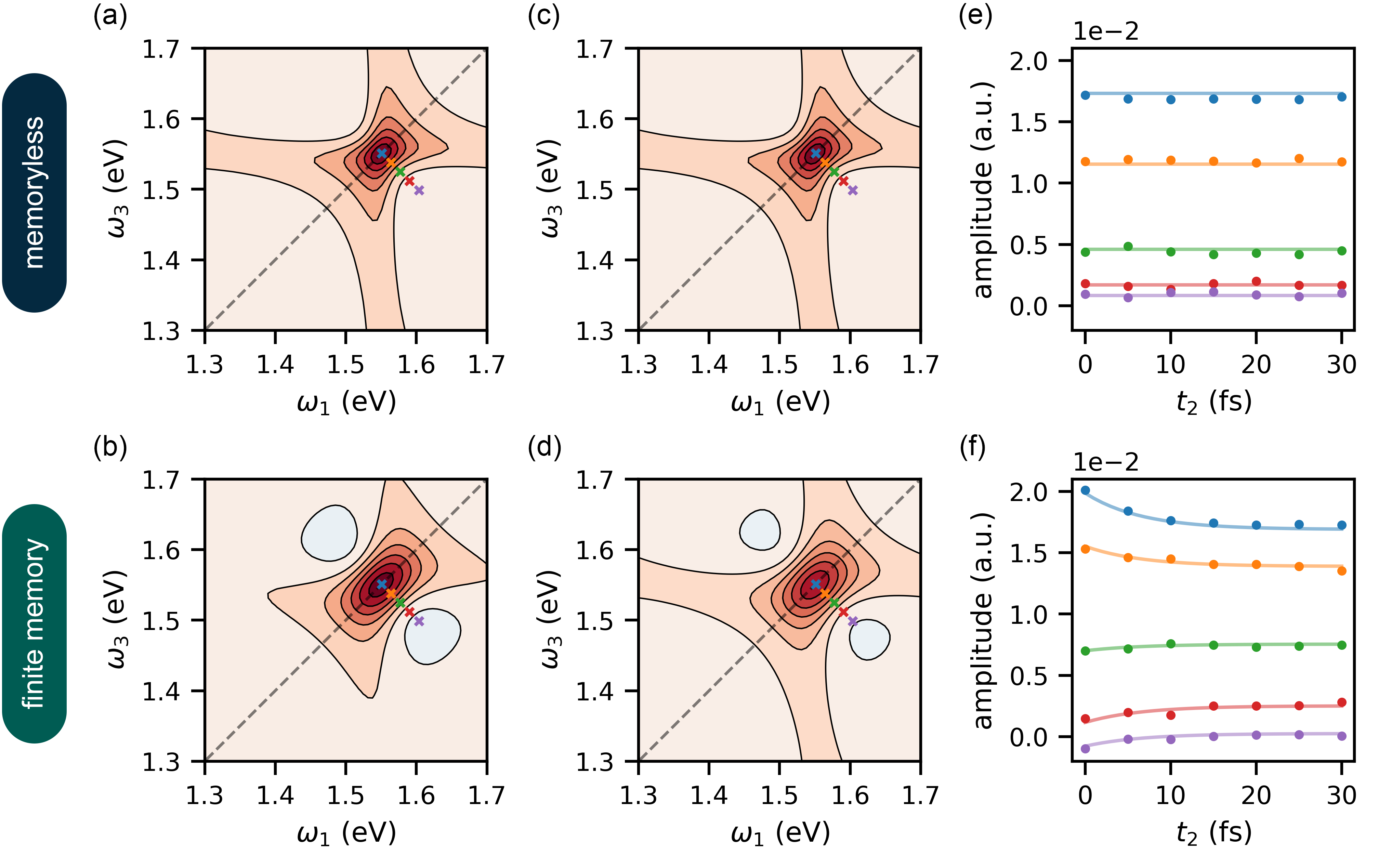}
    \caption{Ground-state bleaching spectra of a single chromophore interacting with a memoryless (a, c, e) and finite-memory (b, d, f) environment. Spectra are collected between $t_2 = 0$ (a, b) and $t_2 = 30\ \si{fs}$ (c, d). No dynamics is observed along the waiting time for the chromophore with a memoryless environment (e), while anti-diagonal spectral diffusion is observed when the environment memory time is finite (f).}
    \label{fig:spectral_diff}
\end{figure*}

In Fig. \ref{fig:spectral_diff}, we show the evolution of the GSB of the single chromophore during the waiting time.
The spectra at $t_2 = 0$ are depicted in Fig. \ref{fig:spectral_diff}a for the memoryless environment, while the finite-memory environment is displayed in Fig. \ref{fig:spectral_diff}b.
The situation at $t_2 = 30\ \si{fs}$ is reported in Fig. 
\ref{fig:spectral_diff}c and Fig. \ref{fig:spectral_diff}d.
As there is only one energy gap in the system, the spectra show a single diagonal peak at $1.55\ \si{eV}$. 
The reported spectra are calculated with the classical simulation of the system-pseudomode evolution, in order to have clean reference spectra (no noise from circuit sampling).
In Fig. 
\ref{fig:spectral_diff}e and Fig. 
\ref{fig:spectral_diff}f, we compare the evolution of the amplitude of the spectra measured at specific points in the 2D maps, indicated by the crosses in the figures, along the line starting at frequency coordinates  $\left( \omega_1, \omega_3 \right) = \left( 1.55\ \si{eV}, 1.55\ \si{eV} \right)$, in blue, to $\left( 1.60\ \si{eV}, 1.50\ \si{eV} \right)$, in violet.
Continuous lines represent the benchmark, while circles are the results of the quantum algorithm simulation, collected with $2 \times 10^4$ sampling shots per circuit.
As can be noted, the inclusion of a finite relaxation rate for the pseudomode influences the signal collected during the waiting time.
In particular, if for the memoryless environment, the spectrum is static, in the case of a pseudomode acting as a memory kernel we observe an anti-diagonal broadening (higher amplitudes are diminished, lower amplitudes are augmented) with a characteristic time of $\hbar \Omega^{-1} = 6.58\ \si{fs}$. This is the spectral diffusion depending on the finite relaxation time of the environment. 

\begin{figure*}
    \centering
    \includegraphics[width=0.8\textwidth]{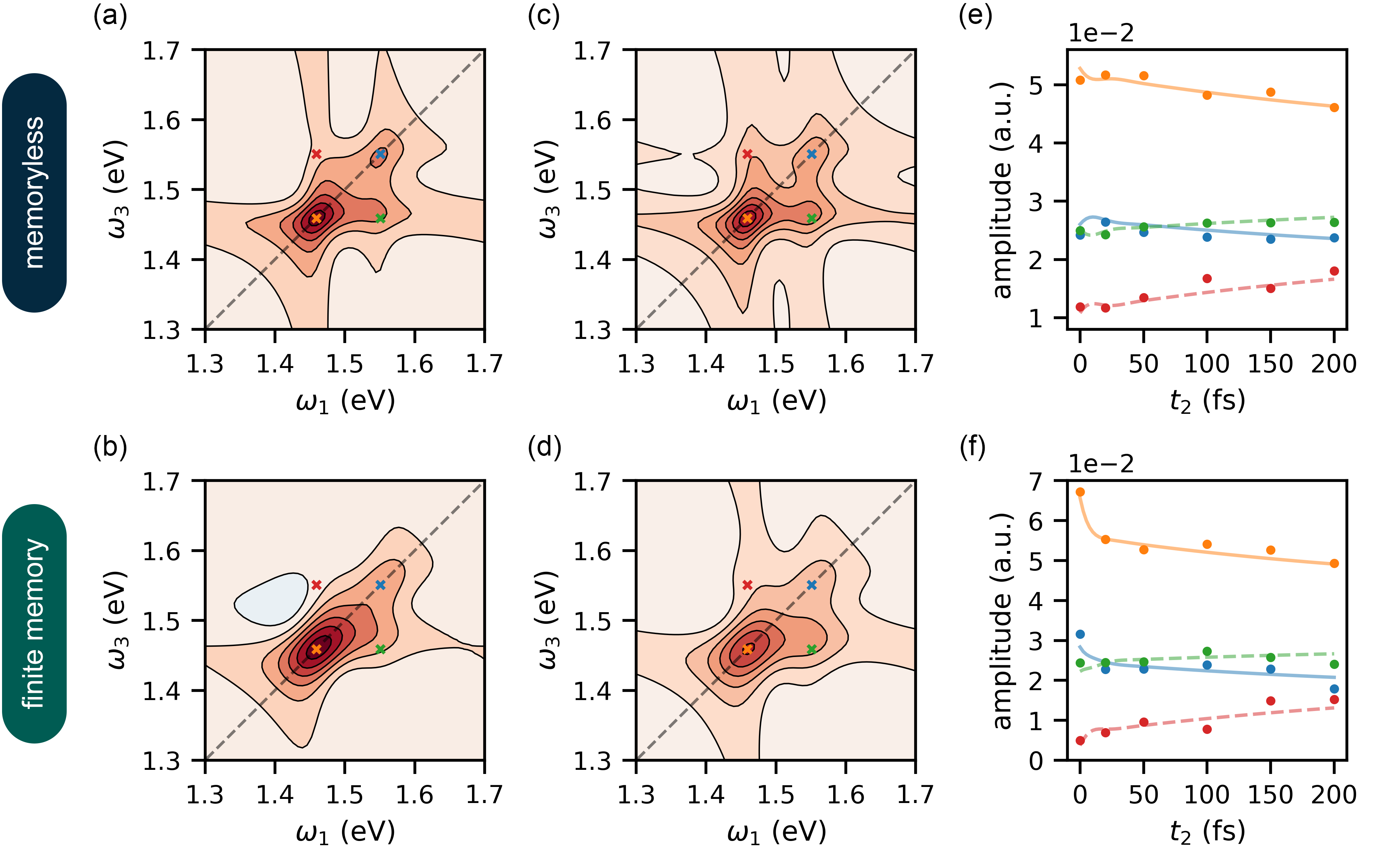}
    \caption{Dynamics along $t_2$ in the third-order rephasing spectra of an excitonic dimer interacting with a memoryless (a, c, e) and finite-memory (b, d, f) environment. Spectra are collected between $t_2 = 0$ (a, b) and $t_2 = 200\ \si{fs}$ (c, d) and show different relaxation dynamics (e,f) of the diagonal (blue and orange) and cross peaks (green and red).}
    \label{fig:dynamics t2}
\end{figure*}

When we pass to the dimer, the situation is slightly more involved.
In Fig. \ref{fig:dynamics t2}, we show the rephasing spectra calculated as the sum of the GSB, SE, and ESA pathways.
Again, Fig. \ref{fig:dynamics t2}a and Fig. \ref{fig:dynamics t2}b shows the spectra at $t_2 = 0$ for the memoryless and finite-memory environment, respectively.
While Fig. \ref{fig:dynamics t2}c and Fig. \ref{fig:dynamics t2}d show the situation at $t_2 = 200\ \si{fs}$.
In Fig. \ref{fig:dynamics t2}e and Fig. \ref{fig:dynamics t2}f, we show the dynamics along the waiting time of the diagonal peaks at $1.55\ \si{eV}$ (blue) and $1.46\ \si{eV}$ (orange), and cross peaks (green and red).
Here, the quantum algorithm (circles) is simulated with $4 \times 10^3$ sampling shots per circuit per pathway.
In Fig. \ref{fig:dynamics t2}e, we can see some initial beatings both in the cross peaks and the diagonal.
The beating of the diagonal peaks can be attributed to nonsecular contributions that are naturally included in the model adopted for the dynamics.
In Fig. \ref{fig:dynamics t2}f, we can recognize the spectral diffusion due to the finite memory time of the pseudomode with the same characteristic time observed for the single chromophore.
This leads to an initial (fast) broadening of the diagonal peaks and covers the initial beatings that are only slightly visible in the dynamics of the cross peaks.
In both environment conditions, the exciton dynamics along $t_2$ continues for a time window longer than the one we simulate, increasing the relative intensity of the cross peaks over the diagonal signal.
%
%%%%%%%%%%%%%%%%%%%%%%%%%%%%%%%%%%%%%%%%%%%%%%%%%%%%%%%%%%%%%%%%%%%%%%%%
%%%%%%%%%%%%%%%%%%%%%%%%%%%%%%%%%%%%%%%%%%%%%%%%%%%%%%%%%%%%%%%%%%%%%%%%
\section{\label{sec:Methods}Methods}
The quantum algorithm for the simulation of a contribution $\alpha$ (eq. \ref{eq:response_function_alpha}) of the time-response function has been previously introduced in Ref. \citenum{Bruschi2024}.
The algorithm consists of a decomposition of the $\alpha$ pathway into site-basis pathways in which single chromophores are excited and de-excited by the light-matter interaction.
The response of each site-basis pathway is then computed via a generalized Hadamard test.
In this work, we integrate the algorithm to include the pseudomode and the collision model which account for the open system dynamics.
The construction of the quantum circuits and their simulation can be performed using our QSExTra module \cite{QSExTra} for Python which exploits tools from the well-known Qutip \cite{Johansson2013} and Qiskit \cite{Qiskit} packages.

This section will address the resource requirements for the embedding of the problem (Section \ref{subsec:Embedding}), as well as the number of circuits to be run and their depth (Section \ref{subsec:Algorithm}).
The structure of the algorithm will be recalled in Section \ref{subsec:Algorithm}.
%%%%%%%%%%%%%%%%%%%%%%%%%%%%%%%%%%%%%%%%%%%%%%%%%%%%%%%%%%%%%%%%%%%%%%%%
\subsection{\label{subsec:Embedding}Embedding in the qubit register}
For the execution of the algorithm, we will need a map from the physical model (exciton system and pseudomodes) to the qubits that compose the quantum register.
Moreover, both the collision model and the algorithm for the response require some ancillary qubits. These are qubits that do not represent physical degrees of freedom but have specific roles in quantum computation.

\subsubsection{Exciton system}
The exciton system is mapped to the quantum register through a one-hot (unary) encoding, where the electronic degrees of freedom of each chromophore are represented by a qubit.
$Q_\text{S} = N$ qubits are needed for an $N$-chromophore aggregate.
The one-hot embedding allows encoding all the possible $2^N$ exciton states from the ground state to the state with $N$-excitations.
The advantage of this mapping is that it allows for the target order of the response to be modified without compromising the overall structure of the algorithm.

\subsubsection{\label{subsubsec:Pseudomodes}Pseudomodes}
For the embedding of the vibrational degrees of freedom, one can take advantage of the Gray code \cite{Bruschi2024}.
By adopting the Gray code, consecutive states of a pseudomode, i.e., states that differ by one occupation number such as $\ket{n}$ and $\ket{n+1}$, are mapped to bitstrings (strings of binary digits) which differ for one bit only.
For example, the first four states, $n = \left\{ 0,1,2,3 \right\}$, are mapped to bitstings $\left\{ 00, 01, 11, 10 \right\}$, respectively.
Then, each resulting bitstring identifies one state of a bunch of qubits.
As noted in Ref. \citenum{Sawaya2020}, this encoding reduces the number of gates needed to decompose operators connecting consecutive states, such as creation and annihilation operators.
The implementation of a (truncated) pseudomode with $d_{ik}$ states requires $\lceil \log_2 d_{ik} \rceil$ qubits of the register.
If all the $W$ pseudomodes have the same number of states $d$, then the pseudomode register will be composed of $Q_\text{P} = N W \lceil \log_2 d_{ik} \rceil$ qubits.

\begin{figure}
    \centering
    \includegraphics[width=0.48\textwidth]{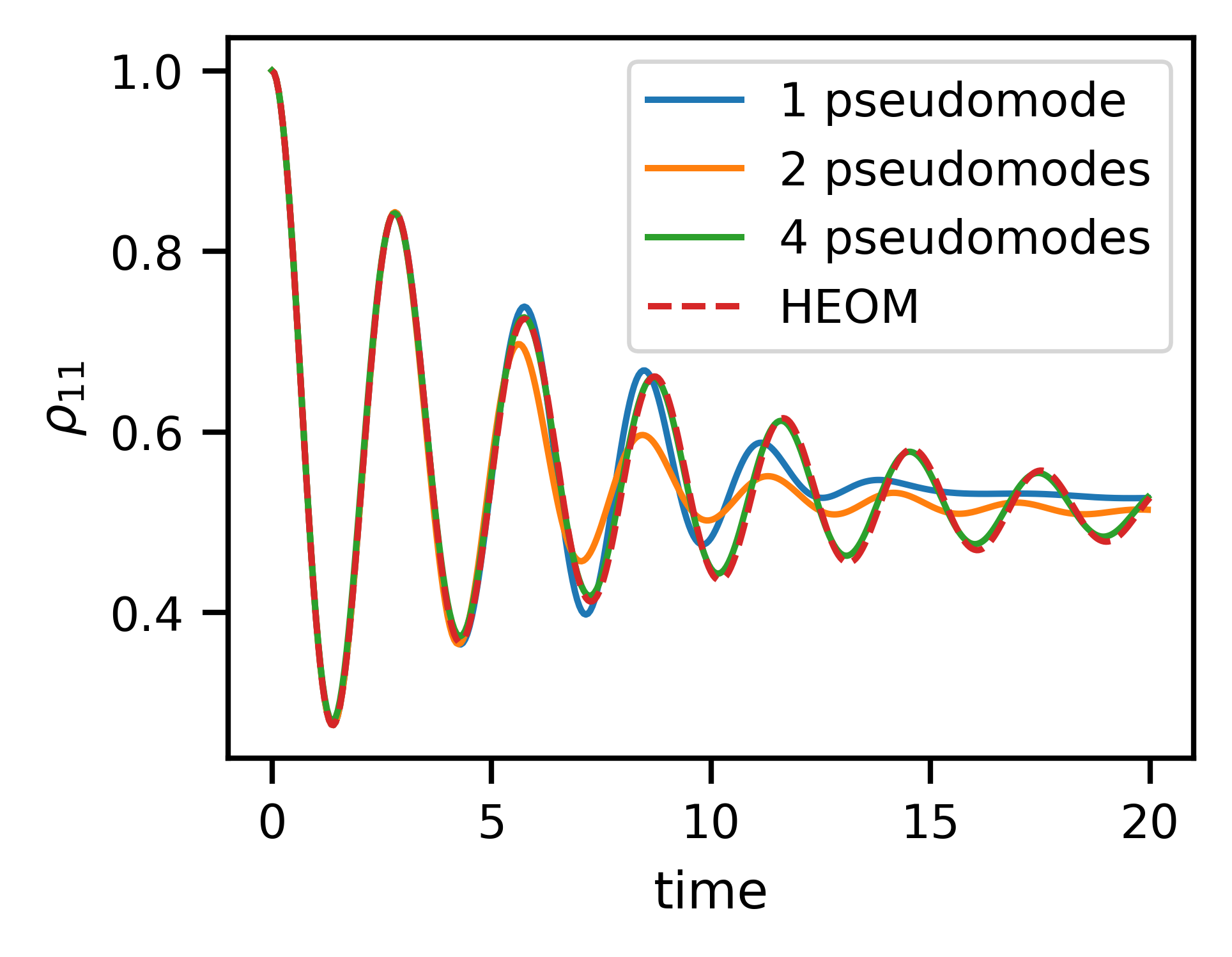}
    \caption{Dynamics of the site-1 population of an excitonic dimer in a highly non-Markovian environment. Simulations are shown for $W'=1$ pseudomode with $d=16$ states (blue line), $W'=2$ pseudomodes with $d=4$ states each (orange line), and $W'=4$ pseudomodes with $d=2$ states each (green line). The number of pseudomodes is intended per chromophore. The result of a HEOM simulation (red line) is reported as a reference. Parameters: $\epsilon_1 = \epsilon_2 = 0$, $J_{12} = 1$, $\omega_k^\text{0} = 0$, $\Gamma = 20 J_{12}$ and $\Omega = 0.1 J_{12}$. Time unit is expressed in terms of the dimension of $J_{12}^{-1}$.}
    \label{fig:more_better}
\end{figure}

When the system-pseudomode coupling is strong, the Hilbert space may require a large number of states, and the truncation may be problematic.
In Figure \ref{fig:more_better}, we show an example in which a single pseudomode with $d = 16$ states (blue line) struggles to reproduce the reference dynamics (HEOM, red line) in the strong coupling regime ($\Gamma/\Omega \gg 1$).
For the same case, using $W'=4$ pseudomodes with $d'=2$ states per pseudomode and a scaled $\Gamma' = \Gamma/W'$ reproduce perfectly the target dynamics (green line).
Further details are given in Appendix \ref{app:more_is_better}.
Notice that in both cases, the dimension of Hilbert space devoted to the pseudomodes is the same ($\mathcal{H}_{1\text{p}} = 16$, $\mathcal{H}_{4\text{p}} = 2^4 = 16$).
In other words, in some regimes, it can be convenient to reformulate the problem to use more pseudomodes with the same frequency and relaxation rate but weaker couplings, that collectively reproduce the same Lorentzian contribution in the spectral function.
If all the $W$ Lorentzians in the spectral function are implemented with $W'$ two-level systems ($d'=2$, i.e, a single qubit), then the pseudomode register is composed of $Q_\text{P} = N W W'$ qubits.

In Section \ref{sec:Results}, simulations of the finite-memory environment are performed with one pseudomode per chromophore truncated to $d=2$ states. The convergence of the results was tested by running simulations of the dynamics with more levels.

\subsubsection{\label{subsubsec:Ancillary qubits}Auxiliary qubits}
The quantum algorithm includes two auxiliary sets of qubits (ancillae), one for the computation of the response function and the other for the collision model which implements the dynamics.

The first set is composed of one ancilla.
This qubit is fundamental for distinguishing bra- and ket-side-oriented interactions between the system and the electric field in the FDs.
Moreover, at the end of a circuit, the state of this ancilla encodes the spectroscopic response of the system, which is extracted by single-qubit measurements.
Therefore, we add to our qubit count a term $Q_\text{A} = 1$. 

\begin{figure}
    \centering
    \includegraphics[width=0.48\textwidth]{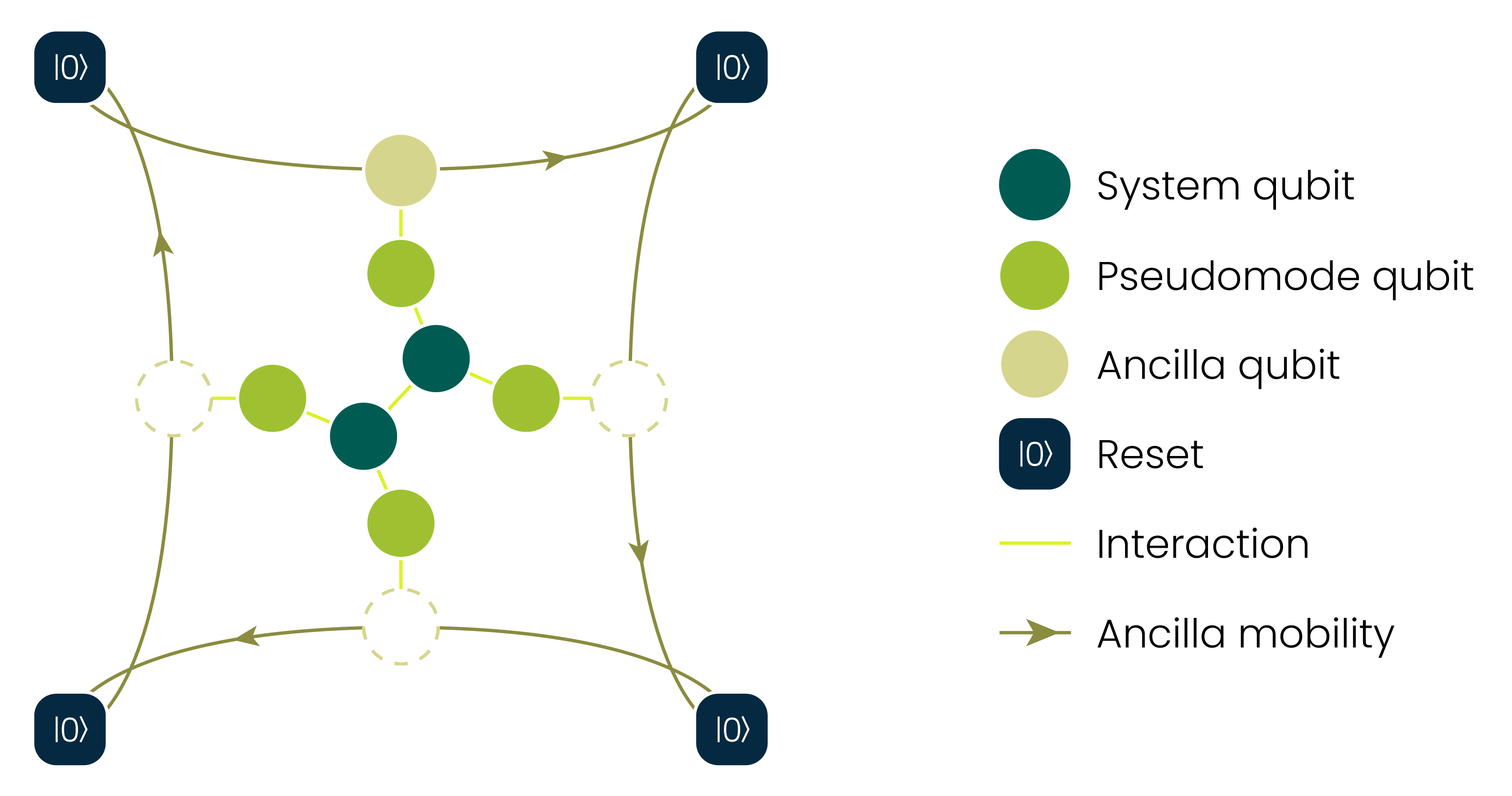}
    \caption{Scheme of a collision model with only one ancilla. The system is composed of two chromophores each one interacting with two truncated pseudomodes with $d=2$ states. The ancilla state is reset after every interaction with a pseudomode qubit to erase memory effects both in time and space (different pseudomodes relax independently).}
    \label{fig:collision_scheme}
\end{figure}

On the other hand, we have the collision model, which requires ancillae as well.
We argue that three different approaches can be used to implement the collision model, which differs in how we interpret the trace on the ancillae in eq. \ref{eq:collision time map} and influences the number of ancillae and reset instructions needed. Let us say that we have to perform $N W$ collisions per time interval (1 collision/pseudomode) and that the dynamics lasts for $t_m/\Delta t$ time intervals. We have the following choices: 
\begin{enumerate}
    \item\label{item:tante ancille}
    Introducing one ancilla for each specific collision and time interval.
    This requires a set of $Q_\text{C} = N W t_m/\Delta t$ qubits.
    However, no reset of the ancillae state is required as the ancillae are used for only one interaction and then discarded.
    \item\label{item:NW ancille}
    A different approach would be to have $Q_\text{C} = N W$ ancillae, each one dedicated to the collision with a specific pseudomode.
    After the collision, the state of each ancilla is reset in order to guarantee the Markovianity of the process.
    This would require $N W t_m/\Delta t$ reset instructions during the dynamics.
    The advantage of the approaches at points \ref{item:tante ancille}-\ref{item:NW ancille} is that different collisions can occur in parallel during the circuit execution and the circuit depth is kept short.
    This is something really valuable when the quantum circuit is executed on a quantum computer.
    However, they require a considerable number of ancillae (large Hilbert space), especially for case 1. Therefore they are not suited for a classical simulation of the quantum algorithm.
    \item\label{item:one ancilla}
    If the number of qubits needs to be minimized at the expense of running a longer circuit, we can use only one ancilla, $Q_\text{C} = 1$, with the caveat of resetting the ancilla state after each collision (Fig. \ref{fig:collision_scheme}).
    This guarantees that different pseudomodes relax independently of the others as the environment is spatially uncorrelated.
    With this method, $N W t_m/\Delta t$ reset instructions are used during the dynamics, but collisions are performed in series.
\end{enumerate}

In simulations of Section \ref{sec:Results}, we always used alternative \ref{item:one ancilla} ($Q_\text{C} = 1$) to keep the Hilbert space as small as possible. However, other choices can become convenient depending on the available resources.
%
%%%%%%%%%%%%%%%%%%%%%%%%%%%%%%%%%%%%%%%%%%%%%%%%%%%%%%%%%%%%%%%%%%%%%%%%
%%%%%%%%%%%%%%%%%%%%%%%%%%%%%%%%%%%%%%%%%%%%%%%%%%%%%%%%%%%%%%%%%%%%%%%%
\subsection{\label{subsec:Algorithm}Algorithm and complexity analysis}
\begin{figure*}
    \centering
    \includegraphics[width=\textwidth]{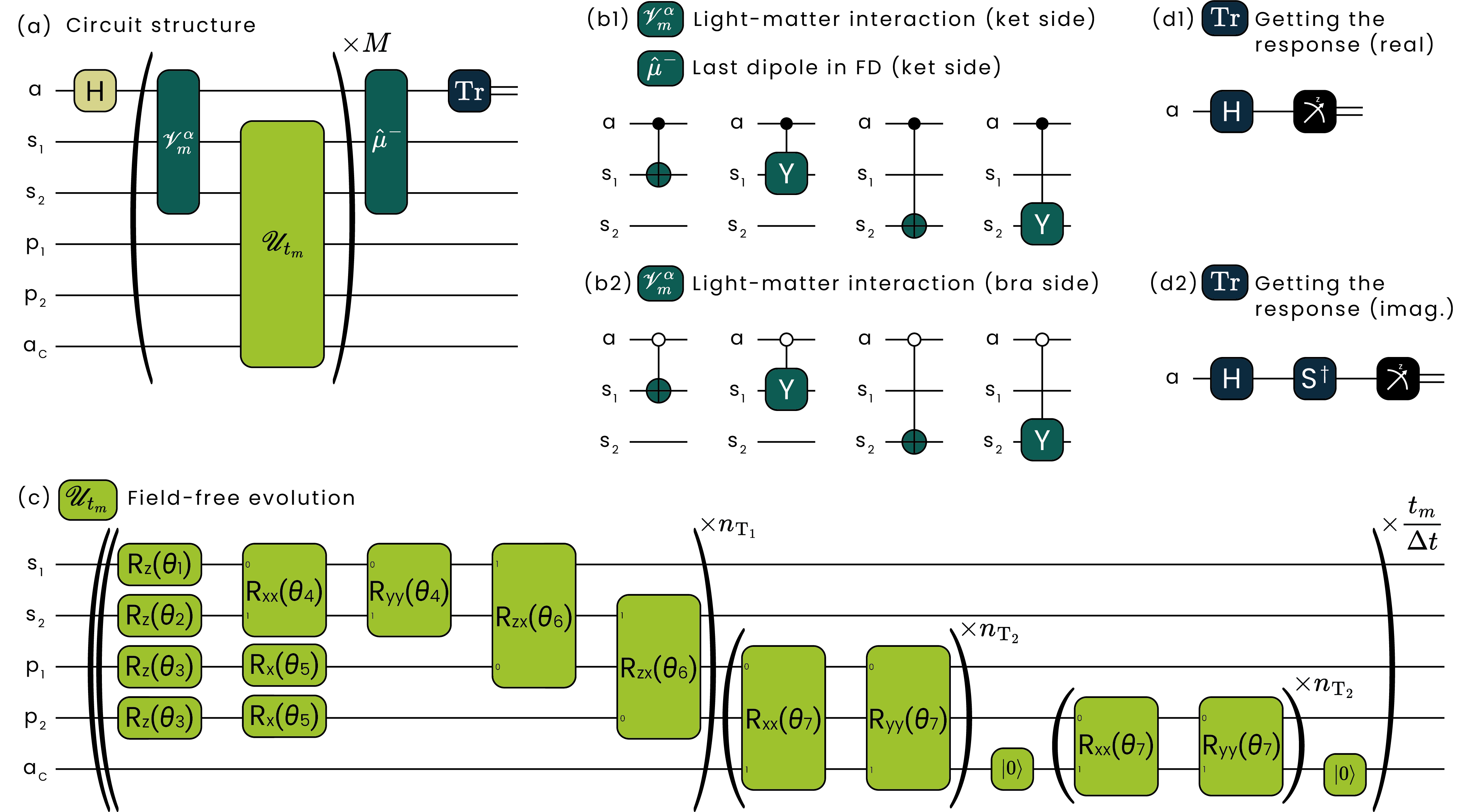}
    \caption{Circuit construction for an interacting dimer. (a) Backbone of the circuit. (b) Gates that implement ket- (b1) and bra- (b2) side light-matter interactions. Close (b1) or open (b2) nodes indicated that the control operation is associated with state 1 or 0 of the control ancilla ``a'', respectively. (c) Trotterized evolution for the system-pseudomode propagator inside the structure of the collision model. Angles $\theta$s are reported in Appendix \ref{app:thetas}. (d) Different measurements of the ancilla state return the real (d1) and imaginary (d2) part of the response function.}
    \label{fig:circuit}
\end{figure*}

Fig. \ref{fig:circuit}a reports the circuit scheme of the algorithm for a dimer system where each chromophore is interacting with a single pseudomode truncated to $d=2$ states.
The ancilla qubit ``a'' is used for the spectroscopic part of the algorithm, ``s$_1$'' and ``s$_2$'' represents the system qubits, ``p$_1$'' and ``p$_2$'' are the pseudomode qubits, one for each chromophore, and ``a$_\text{c}$'' denotes the ancilla involved in the collision model.
The same register has been used to obtain the results for the excitonic dimer with pseudomodes in Section \ref{subsec:Nonlinear spectra}.\\
At the beginning of the circuit, a Hadamard gate prepares an ancilla qubit in the superposition state $\ket{\psi_\text{a}} = \left( \ket{0_\text{a}} + \ket{1_\text{a}} \right) / \sqrt{2}$.
According to an algorithmic vision of FDs \cite{Bruschi2024}, we then enter an interaction-evolution loop of length $M$ (the order of the desired response). In the following, we derive the scaling of the algorithm with $M$ and $N$ (the number of chromophores). This is done by counting the required two-qubit gates (CNOT count) and evaluating the circuit depths.
%
%%%%%%%%%%%%%%%%%%%%%%%%%%%%%%%%%%%%%%%%%%%%%%%%%%%%%%%%%%%%%%%%%%%%%%%%
\subsubsection{\label{subsubsec:qc_interaction}Interaction $\mathcal{V}_m^\alpha$: Encoding and scaling}
To reproduce the effect of the light-matter interaction, we need (\ref{item:gate_encoding}) a gate encoding for the dipole moment operators and (\ref{item:controlled_gates}) a way to distinguish between the ket- and bra-side interactions.
In the following, we analyze the circuit construction for the light-matter interaction block step by step.
Notice that, the signal emission in a FD corresponds to the application of a $(M+1)$-th dipole moment ($\hat{\mu}^-$) on the ket side.
Formally, this has the same structure as a light-matter interaction term and is thus treated as a ket-side de-excitation interaction.
\begin{enumerate}
    \item\label{item:gate_encoding}
    For the quantum algorithm, we associate the excitations (incoming arrows in the FD) on both the ket and bra with $\hat{\mu}^+$, while de-excitations (outgoing arrows in the FD) are associated with $\hat{\mu}^-$.
    To encode the dipole moments $\hat{\mu}^\pm$, we have to transform them into unitary operators.
    We do so by taking advantage of eq. \ref{eq:dipole_operators}, where dipole moments are written as a sum of local contributions.
    By substituting explicitly such a decomposition into eq. \ref{eq:response_function_alpha} and moving the summation outside of the correlation function, it is evident that the response function of a pathway $\alpha$ can be interpreted as the sum of $N^{M+1}$ terms coming from the combination of light-matter interactions with single chromophores.
    We name each of these terms as a ``site-basis pathway''.
    However, to recover the unitarity of the ladder operators used in site-basis pathways, we need to decompose them further in terms of Pauli-X and Pauli-Y as $\hat{\sigma}^\pm = \left( \hat{\sigma}^x - i\hat{\sigma}^y \right)$.
    Therefore, the response function is eventually decomposed in $\left(2N\right)^{M+1}$ correlation functions in which light-matter interaction operators are local (act on a single chromophore) and unitary.
    Each of these correlation functions is implemented as an independent quantum circuit.
    Finally, at the end of the computation, the outcome of all circuits is summed up and the response is obtained.
    As discussed in Ref. \citenum{Bruschi2024}, if the number of excitations is conserved during the delay times, we can reduce the number of circuits up to $2^{M-2}N^{M+1}$ for the case of an aggregate and $N^{M+1}$ for a single chromophore.
    \item\label{item:controlled_gates}
    To distinguish between the ket- and bra-side interactions, we exploit the superposition state of the ancilla created at the beginning of the circuit and controlled gates.
    A controlled gate applies an operation to a target qubit only if the control qubit is in a certain state.
    In particular, we frame the algorithm so that ket-side interactions are controlled by state $\ket{1_\text{a}}$ of the ancilla, while bra-side interactions happen when $\ket{0_\text{a}}$.
    By virtue of the decomposition introduced in the point above, the resulting pool of controlled gates consists of the controlled Pauli-X (CX or CNOT) and controlled Pauli-Y (CY) gates.
    Figs. \ref{fig:circuit}b1 and \ref{fig:circuit}b2 represent the gates needed for a dimer system for ket- and bra-side interactions, respectively.
\end{enumerate}
Therefore, inside a quantum circuit, the implementation of the light-matter interaction contributes to the CNOT count and depth with $M+1 = \mathcal{O}\left( M \right)$ gates.
%%%%%%%%%%%%%%%%%%%%%%%%%%%%%%%%%%%%%%%%%%%%%%%%%%%%%%%%%%%%%%%%%%%%%%%%
\subsubsection{Evolution $\mathcal{U}_{t_m}$: Encoding and scaling}
During the delay times, the system qubits evolve under the action of the dynamical map $\mathcal{U}_{t_m}$.
Let us take Fig. \ref{fig:circuit}c as a reference.
We build a quantum circuit that evolves the system for a time step $\Delta t$ according to the collision model (see eq. \ref{eq:collision time map}).
The dynamics for a time $t_m$ is obtained by repeating the circuit for $t_m / \Delta t$ times.
The circuit is composed of two blocks.
The first block consists of the propagator for the Hamiltonian $\hat{H}$, eq. \ref{eq:Hamiltonian propagator}.
The second block includes the collisions between the pseudomodes and the ancilla, eq. \ref{eq:collision propagator}, and the reset of the ancilla state.
In both blocks, propagators can be implemented in various ways \cite{Childs2018}, however, we use standard first-order Trotter decomposition.

The Trotterized evolution due to the system Hamiltonian requires $\mathcal{O}\left( N^2 \right)$ CNOT gates for simulating the interaction between the chromophores in a fully connected aggregate (short- and long-range interactions), while the correspondent depth scales linearly as $\mathcal{O}(N)$ \cite{Bruschi2024, Gallina2024}.
For pseudomodes truncated at $d>2$, the Grey code makes it easy to reduce their contribution in terms of Pauli strings \cite{Bruschi2024, Sawaya2020}.
Then, a CNOT-staircase decomposition \cite{Sawaya2020, Bruschi2024, Gallina2024} can be used to implement the pseudomode self-evolution, and the chromophore-pseudomode and pseudomode-ancilla interactions.
This requires $\mathcal{O}\left( NWd \log^2 d \right)$ CNOT gates per Trotter step \cite{Bruschi2024}.
However, in light of what we have discussed in Section \ref{subsubsec:Pseudomodes}, sometimes it can be convenient to include multiple identical pseudomodes truncated to $d=2$ states.
In this case, creation and annihilation operators coincide with qubit ladder operators ($\hat{a} = \hat{\sigma}^-$ and $\hat{a}^\dagger = \hat{\sigma}^+$).
All the gates which include the pseudomodes become trivial.
In particular, pseudomodes interactions are represented by two-qubit gates, and the CNOT scaling of all gates that acts on the pseudomodes increases as $\mathcal{O}\left( NWW' \right)$ CNOTs per Trotter step.

The time step $\Delta t$ should be chosen by performing the dynamics of the system for different time steps and checking the convergence.
If $\Delta t$ is kept small enough, for instance when $\Omega$ is comparable to the system timescale, Trotterization can be implemented with a single Trotter step ($n_{\text{T}_1} = n_{\text{T}_2} = 1$).
Anyway, when pseudomodes are truncated to $d=2$ states, one can always set $n_{\text{T}_2} = 1$.

%%%%%%%%%%%%%%%%%%%%%%%%%%%%%%%%%%%%%%%%%%%%%%%%%%%%%%%%%%%%%%%%%%%%%%%%
\subsubsection{Measuring and retrieving the  response function}
When two-qubit gates are applied according to the description provided in Section \ref{subsubsec:qc_interaction}, the ancilla qubit (``a'') becomes entangled with the system state in a way that the contribution to the response function can be retrieved by measuring the expectation value of specific operators of the ancilla. 
In particular, $\langle \hat{\sigma}^x_\text{a} \rangle$ (Fig. \ref{fig:circuit}d1) contributes to the real part of the response, while $\langle \hat{\sigma}^y_\text{a} \rangle $ (Fig. \ref{fig:circuit}d2) contributes to the imaginary part.

Once the responses of the site-basis pathways have been collated for all the relevant combinations of CX and CY gates, the complete response function of the aggregate can be obtained by summing these contributions and weighting them according to the amplitudes of the dipole moments involved.

%%%%%%%%%%%%%%%%%%%%%%%%%%%%%%%%%%%%%%%%%%%%%%%%%%%%%%%%%%%%%%%%%%%%%%%%
%%%%%%%%%%%%%%%%%%%%%%%%%%%%%%%%%%%%%%%%%%%%%%%%%%%%%%%%%%%%%%%%%%%%%%%%
\section{Conclusions and Outlook}
In this paper, we introduce a theoretical framework allowing the quantum simulation of the linear and nonlinear optical response of Frenkel excitonic systems interacting with an arbitrary bosonic environment, without limiting assumptions on the strength of the exciton-vibrational interaction. The workflow integrates (i) a Markovian embedding through the definition of an equivalent environment in terms of pseudomodes, (ii) a collision model of the resulting dynamics and (iii) a decomposition of the optical response function into site-basis pathways. We demonstrate that this theoretical framework can be fully implemented in a digital quantum computer, therefore harnessing the so-called quantum advantage for the simulation of quantum many-body dynamics. The advantage can be understood both at the level of storage and algorithmic complexity. Indeed, the total number of qubits required in the simulation of the quantum algorithm is $Q = \sum_{x=\{\text{S,P,C,A}\}} Q_x = N \left(1 + W \lceil \log_2 d \rceil \right) + 2$, where $N$ is the number of chromophores in the aggregate and $W$ the number of truncated pseudomodes (per chromophore) with $d$ levels. Encoding the density matrix of the same exciton-pseudomode system in a classical setting requires resources scaling exponentially with $N$ and $W$ \cite{Bruschi2024}. Consequently, the same scaling is required for the evolution in a time step, as it requires the multiplication between the evolution matrix and the quantum state. On the other hand, the overall number of gates per Trotter step in the quantum circuit scales (at most) quadratically with the number of chromophores, linearly with the number of pseudomodes per chromophore, and quasilinearly with the number of states per pseudomode.

On the other hand, the overall performance of the quantum algorithm is hindered by the number of site-basis pathways that must be evaluated, which increases with the power of the order of the response ($M+1$). Even if the order of the nonlinear response is commonly $M=3$, or $M=5$ in the latest developments \cite{Luttig2023}, two strategies can be used to mitigate the number of these pathways. On the one hand, one can exclude certain site-basis pathways based on the localization length of the excitons. It is well known that in extended chromophore systems, excitons are spread over a limited number of chromophores. Therefore, pathways which involve the excitation of one site and subsequent de-excitation of another site located on a distant portion of the aggregate can be neglected as their contribution will be minimal.
This pathway selection can be done a priori without modifying the quantum algorithm.
An alternative approach would be to work on the exciton basis, whereby pathways can be neglected based on the brightness and darkness of the excitonic states. This involves diagonalizing the system Hamiltonian and reformulating the quantum algorithm on the new basis.

Some final remarks on the Markovian embedding with pseudomodes are in order. The underlying theory \cite{Tamascelli2018, Pleasance2021} ensures the system dynamics is reproduced exactly by the pseudomode embedding when the correlation function of the original bosonic bath can be written as a finite sum of complex exponentials. In other words, the fitting of an arbitrary environment spectral function in terms of Lorentians may be a source of error. We refer the interested reader to Refs.\citenum{Mascherpa2020, Pleasance2021, Lambert2019} for a detailed discussion of this issue. In practice, the target spectral densities for excitonic systems derive from experimental results \cite{Gustin2023} or atomistic simulations \cite{Valleau2012} and therefore they are already affected by an intrinsic uncertainty. In this context, one can always improve the pseudomode embedding to the extent that the fitting error becomes irrelevant for the observables of interest \cite{Leppakangas2023, Mascherpa2020}. 

To conclude, we point out interesting issues related to this work calling for further investigations. A question is whether there exists an alternative Markovian embedding schemes that optimize the quantum resources used in the simulation of exciton dynamics. For example, Ref. \citenum{Mascherpa2020} considers a set of interacting surrogate oscillators rather than independent pseudomodes, while the reaction coordinate model \cite{Iles-Smith2016} maps a general environment into a single-mode reaction coordinate and a residual Markovian bath. Studying how these different embedding schemes reflect on the width and depth of quantum circuits may lead to further resource optimization. Notice that the requirement of a quantum implementation poses specific constraints on the embedding model. For example, the generalization of pseudomodes to non-hermitian coupling Hamiltonian, discussed in Refs. \citenum{Pleasance2020,Pleasance2021}, is not practical in the context of a quantum circuit where the dynamics is implemented through unitary gates.
\\
On the other hand, the decomposition of the environment correlation function into quantum and classical components, as proposed in \cite{Luo2023, Cirio2024}, may be translated into quantum propagation algorithms via parametrized random quantum circuits as those used in Ref. \citenum{Gallina2024}. Testing the combination of Markovian embedding and stochastic Hamiltonian in the quantum simulation of spectroscopy is a promising direction to improve the efficiency or the accuracy of the proposed workflow.
\\
Finally, the use of quantum phase estimation (QPE) for the simulation of nonlinear spectroscopic was recently explored in Ref. \cite{Loaiza2024}. Understanding whether non-unitary effects due to the system-environment interaction can be included in the framework of QPE, beyond the use of window functions, remains an open question worthy of further consideration.
%%%%%%%%%%%%%%%%%%%%%%%%%%%%%%%%%%%%%%%%%%%%%%%%%%%%%%%%%%%%%%%%%%%%%%%%
%%%%%%%%%%%%%%%%%%%%%%%%%%%%%%%%%%%%%%%%%%%%%%%%%%%%%%%%%%%%%%%%%%%%%%%%
\begin{acknowledgments}
This work is (partially) supported by ICSC – Centro Nazionale di Ricerca in High Performance Computing, Big Data and Quantum Computing, funded by European Union – NextGenerationEU and by the Department of Chemical Sciences (DiSC) and the University of Padova within the Project QA-CHEM (P-DiSC No. 04BIRD2021-UNIPD)

The authors acknowledge the C3P facility of the Department of Chemistry of the University of Padova for the computing resources for the simulations. F.G. and B.F. acknowledge the CINECA award under the ISCRA initiative, for the availability of high-performance computing resources and support.
\end{acknowledgments}
%%%%%%%%%%%%%%%%%%%%%%%%%%%%%%%%%%%%%%%%%%%%%%%%%%%%%%%%%%%%%%%%%%%%%%%%
%%%%%%%%%%%%%%%%%%%%%%%%%%%%%%%%%%%%%%%%%%%%%%%%%%%%%%%%%%%%%%%%%%%%%%%%
\appendix
\section{\label{app:collision_to_master_equation}From collision model to Lindblad master equation} 
A collision step evolves the system state as
\begin{equation} \label{app_eq:propagation}
    \rho \left( t + \delta t \right) =
    \Tr_\text{C}
    \left\{
    \hat{U}^\text{C} \left( \delta t \right)
    \hat{U} \left( \delta t \right)
    %\left(
    \rho (t) \otimes \rho_\text{C}^0
    %\right)
    \hat{U}^\dagger \left( \delta t \right)
    \hat{U}^{\text{C} \dagger} \left( \delta t \right)
    \right\}
\end{equation}
where the state of the ancillae before the collision is the product state $\rho_\text{C}^0 = \bigotimes_{i,k} \ketbra{0_\text{C} (ik)}{0_\text{C}(ik)}$ and the propagator of the collision is $\hat{U}^\text{C} = \exp{-i \delta t \hat{H}^\text{C} \left( \delta t \right)}$ with $\hat{H}^\text{C} \left( \delta t \right) = \sum_{i=1}^N \sum_{k=1}^{W} \hat{H}_{ik}^\text{C} \left( \delta t \right)$.
Notice that Hamiltonians $\hat{H}_{ik}^\text{C}$ always commute since they act on different spaces.

For small time intervals, we can expand both propagators at first order in $\delta t$ (remember that $\hat{H}^\text{C} \left( \delta t \right) \propto \delta t^{-1/2}$), so that
\begin{equation}
    \hat{U} \left( \delta t \right) =
    1
    - i \hat{H} \delta t
\end{equation}
and
\begin{equation}
    \hat{U}^\text{C} \left( \delta t \right) =
    1
    - i \hat{H}^\text{C} \left( \delta t \right) \delta t
    - \frac{1}{2} \left( \hat{H}^\text{C} \left( \delta t \right) \right)^2 \delta t^2.
\end{equation}
By inserting these expressions in eq. \ref{app_eq:propagation} and keeping terms up to the linear order in $\delta t$, we get
\begin{equation} \label{app_eq:step1}
\begin{split}
    \rho \left( t + \delta t \right) &=
    \rho \left( t \right) \\
    & - i \left[ \hat{H}, \rho \left( t \right) \right] \delta t \\
    & - i \Tr_\text{C} \left\{ \left[ \hat{H}^\text{C} \left( \delta t \right), \left( \rho (t) \otimes \rho_\text{C}^0 \right) \right] \right\} \delta t \\
    & + \Tr_\text{C} \left\{ \hat{H}^\text{C} \left( \delta t \right) \left( \rho (t) \otimes \rho_\text{C}^0 \right) \hat{H}^\text{C} \left( \delta t \right) \right\} \delta t^2 \\
    & - \frac{1}{2} \Tr_\text{C} \left\{ \left[ \left( \hat{H}^\text{C} \left( \delta t \right) \right)^2, \left( \rho (t) \otimes \rho_\text{C}^0 \right) \right]_+ \right\} \delta t^2.
\end{split}
\end{equation}
Since $\Tr_\text{C} \left\{ \hat{\sigma}_\text{C}^\pm (ik) \rho_\text{C}^0 \right\} = 0$, it is easy to show that
\begin{equation}
    \Tr_\text{C}
    \left\{
    \hat{H}_{ik}^\text{C} \left( \delta t \right)
    \left( \rho (t) \otimes \rho_\text{C}^0 \right)
    \right\} =
    0.
\end{equation}
In the same way, from
\begin{align}
    \Tr_\text{C} \left\{ \hat{\sigma}_\text{C}^+ (ik) \hat{\sigma}_\text{C}^+ (i'k') \rho_\text{C}^0 \right\} &= 0, \\
    \Tr_\text{C} \left\{ \hat{\sigma}_\text{C}^- (ik) \hat{\sigma}_\text{C}^- (i'k') \rho_\text{C}^0 \right\} &= 0, \\
    \Tr_\text{C} \left\{ \hat{\sigma}_\text{C}^+ (ik) \hat{\sigma}_\text{C}^- (i'k') \rho_\text{C}^0 \right\} &= 0, \\
    \Tr_\text{C} \left\{ \hat{\sigma}_\text{C}^- (ik) \hat{\sigma}_\text{C}^+ (i'k') \rho_\text{C}^0 \right\} &= \delta_{ii'} \delta_{kk'},  
\end{align}
where $\delta_{ab}$ denotes the Kronecker delta, it follows that
\begin{equation}
    \Tr_\text{C}
    \left\{
    \hat{H}_{ik}^\text{C} \left( \delta t \right)
    \left( \rho (t) \otimes \rho_\text{C}^0 \right)
    \hat{H}_{i'k'}^\text{C} \left( \delta t \right)
    \right\} =
    \frac{2 \Omega_k}{\delta t} 
    \hat{a}_{ik} \rho (t) \hat{a}_{ik}^\dagger
    \delta_{ii'} \delta_{kk'},
\end{equation}
\begin{equation}
    \Tr_\text{C}
    \left\{
    \hat{H}_{ik}^\text{C} \left( \delta t \right)
    \hat{H}_{i'k'}^\text{C} \left( \delta t \right)
    \left( \rho (t) \otimes \rho_\text{C}^0 \right)
    \right\} =
    \frac{2 \Omega_k}{\delta t} 
    \hat{a}_{ik}^\dagger \hat{a}_{ik} \rho (t)
    \delta_{ii'} \delta_{kk'},
\end{equation}
\begin{equation}
    \Tr_\text{C}
    \left\{
    \left( \rho (t) \otimes \rho_\text{C}^0 \right)
    \hat{H}_{ik}^\text{C} \left( \delta t \right)
    \hat{H}_{i'k'}^\text{C} \left( \delta t \right)
    \right\} =
    \frac{2 \Omega_k}{\delta t} 
    \rho (t) \hat{a}_{ik}^\dagger \hat{a}_{ik}
    \delta_{ii'} \delta_{kk'}.
\end{equation}
Using the results above, we can express eq. \ref{app_eq:step1} as
\begin{equation} \label{app_eq:step2}
    \begin{split}
        \rho \left( t + \delta t \right) = &
        \rho \left( t \right) + \\
        & - i \left[ \hat{H}, \rho \left( t \right) \right] \delta t + \\
        & + 2 \delta t \sum_{i=1}^N \sum_{k=1}^{W} \Omega_k \left( \hat{a}_{ik} \rho (t) \hat{a}_{ik}^\dagger - \frac{1}{2} \left[ \hat{a}_{ik}^\dagger \hat{a}_{ik}, \rho (t) \right]_+ \right).
    \end{split}
\end{equation}
Now, in the limit of $\delta t \to 0$,
\begin{equation}
    \lim_{\delta t \to 0}\frac{\rho \left( t + \delta t \right) - \rho \left( t \right)}{\delta t} = \frac{d \rho (t)}{dt}
\end{equation}
and the collision model approaches the target master equation
\begin{equation}
\begin{split}
    \frac{d \rho (t)}{dt} = & -i \left[ \hat{H}, \rho (t) \right] \\
    & + 2 \sum_{i=1}^N \sum_{k=1}^{W} \Omega_k \left( \hat{a}_{ik} \rho (t) \hat{a}_{ik}^\dagger - \frac{1}{2} \left[ \hat{a}_{ik}^\dagger \hat{a}_{ik}, \rho (t) \right]_+ \right).
\end{split}
\end{equation}
%%%%%%%%%%%%%%%%%%%%%%%%%%%%%%%%%%%%%%%%%%%%%%%%%%%%%%%%%%%%%%%%%%%%%%%%
%%%%%%%%%%%%%%%%%%%%%%%%%%%%%%%%%%%%%%%%%%%%%%%%%%%%%%%%%%%%%%%%%%%%%%%%
\section{\label{app:relax_and_trap}Including exciton relaxation and trapping within the collision model}
The flexibility of the collision model makes it easy to account for additional relaxation sources \cite{Cattaneo2021}.
Relevant examples in light-harvesting complexes are natural thermal relaxation and exciton transfer to a trap state, which are essential to estimate the transport efficiency to the reaction center.
Other sources can be, for example, the presence of a sink or a perturbative electric field with the right wavelength.
For simplicity, here we assume that the electronic energy gap is large enough so that, at thermal equilibrium, the population is in the ground state only.
The dissipator can be expressed as
\begin{equation} \label{eq:dissipator_relaxation}
    \mathcal{D}^\text{R} \left[ \rho (t) \right] =
    \sum_{i=1}^N
    \gamma_{\text{R} i}
    \left(
    \hat{\sigma}_i^- \rho (t) \hat{\sigma}_i^+
    - \frac{1}{2} \left[ \hat{\sigma}_i^+ \hat{\sigma}_i^-, \rho (t) \right]_+
    \right),
\end{equation}
where the effect is that of a de-excitation of the chromophoric network during time.
The collision with an ancilla in the initial state $\ket{0_\text{R}}$ that reproduces the dissipator in eq. \ref{eq:dissipator_relaxation} is dictated by Hamiltonian
\begin{equation}
    \hat{H}^\text{R} = \sum_{i=1}^N \hat{H}_i^\text{R} =
    \sum_{i=1}^N \sqrt{\frac{\gamma_{\text{R} i}}{\delta t}}
    \left(
    \hat{\sigma}_i^+ \hat{\sigma}_\text{R}^- (i) +
    \hat{\sigma}_i^- \hat{\sigma}_\text{R}^+ (i)
    \right),
\end{equation}

Then, we assume the transfer from a target site $j$ to a reaction center.
To do so, we enlarge the Hilbert space including an extra qubit which encodes the vacuum state $\ket{0_\text{rc}}$ and the occupied states $\ket{1_\text{rc}}$ of the reaction center.
The density matrix of the system plus reaction center is indicated by $\rho(t) \otimes \rho_\text{rc} (t)$.
The dissipator of the process is
\begin{equation} \label{eq:dissipator_trap}
\begin{split}
    \mathcal{D}^\text{T} \left[ \rho (t) \right] =
    \gamma_\text{T}
    \bigg( &
    \hat{\sigma}_j^- \hat{\sigma}_\text{rc}^+ \left( \rho(t) \otimes \rho_\text{rc} (t) \right) \hat{\sigma}_j^+ \hat{\sigma}_\text{rc}^-
    +\\
    & - \frac{1}{2} \left[ \hat{\sigma}_j^+ \hat{\sigma}_j^- \hat{\sigma}_\text{rc}^- \hat{\sigma}_\text{rc}^+, \rho(t) \otimes \rho_\text{rc} (t) \right]_+
    \bigg),
\end{split}
\end{equation}
and the relative collision Hamiltonian with an ancilla initialized in state $\ket{0_\text{T}}$ is
\begin{equation}
    \hat{H}^\text{T} =
    \sqrt{\frac{\gamma_\text{T}}{\delta t}}
    \left(
    \hat{\sigma}_j^+ \hat{\sigma}_\text{rc}^- \hat{\sigma}_\text{T}^- +
    \hat{\sigma}_j^- \hat{\sigma}_\text{rc}^+ \hat{\sigma}_\text{T}^+
    \right).
\end{equation}
%%%%%%%%%%%%%%%%%%%%%%%%%%%%%%%%%%%%%%%%%%%%%%%%%%%%%%%%%%%%%%%%%%%%%%%%
%%%%%%%%%%%%%%%%%%%%%%%%%%%%%%%%%%%%%%%%%%%%%%%%%%%%%%%%%%%%%%%%%%%%%%%%
\section{\label{app:more_is_better}Decomposition of a highly non-Markovian  pseudomode}
\begin{figure}
    \centering
    \includegraphics[width=0.48\textwidth]{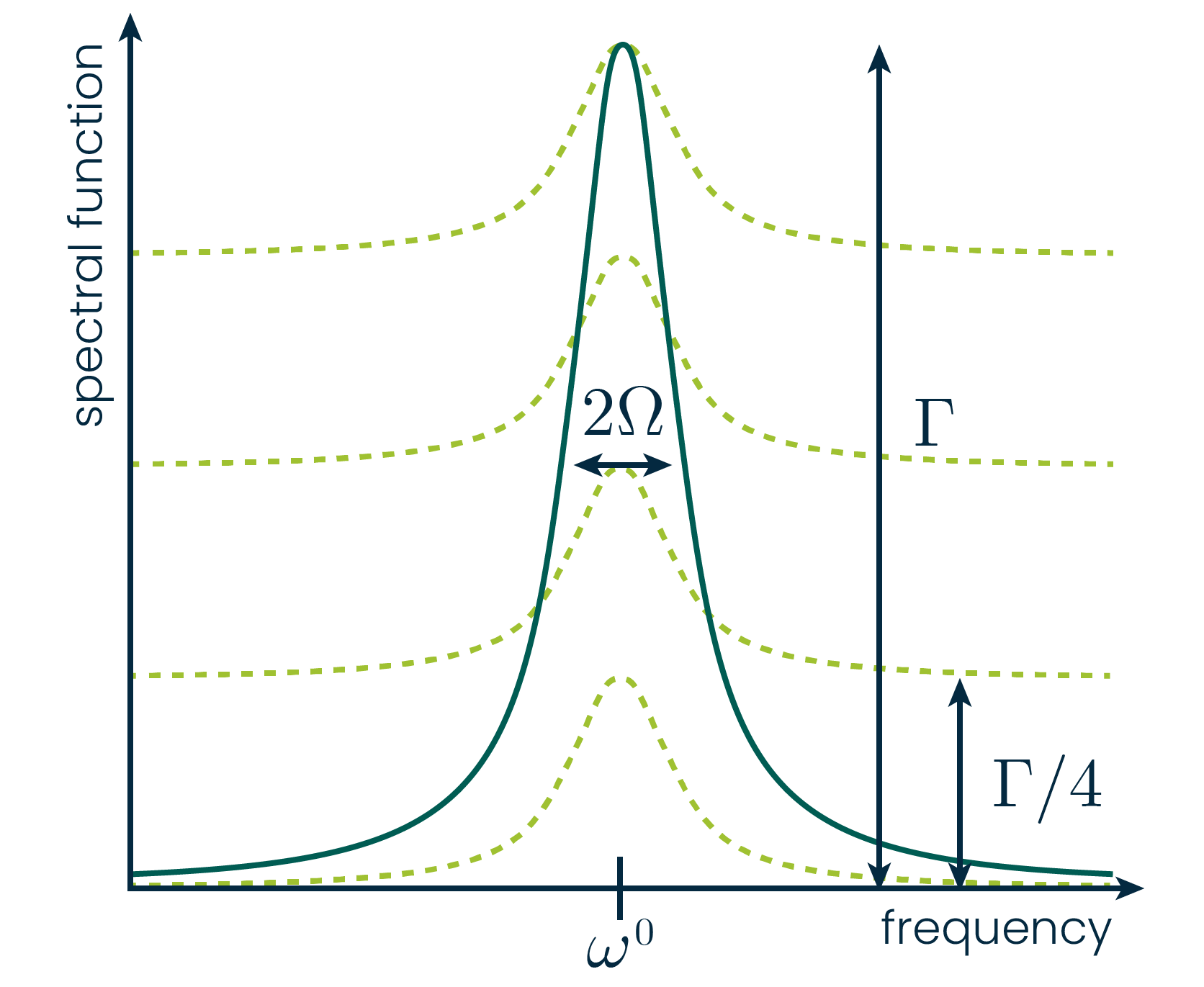}
    \caption{A Lorentzian (spectral) function (solid line) with width $\Omega$ and amplitude $\Gamma$ is decomposed as the sum of 4 Lorentzians with the same width $\Omega$ but scaled amplitude $\Gamma/4$. The center of the Lorentzian ($\omega^0$) remains unchanged.}
    \label{fig:Lorentzian_sum}
\end{figure}

In the case of a highly non-Markovian pseudomode, i.e., when $\Omega_k$ is small and/or $\Gamma_k$ is large, the reproduction of the spectral function may require the inclusion of several states ($d$) of the harmonic oscillator due to the rapid saturation of its active space.
In general, saturation is strictly correlated to the $\Gamma_k/\Omega_k$ ratio.
We observed that a useful method to enhance the performance of the simulation without increasing the computational costs was to split the pseudomode into multiple pseudomodes while maintaining the total Hilbert space dimension.
In fact, a certain Lorentzian with amplitude $\Gamma$ can be easily decomposed as the sum of $W'$ Lorentzians with amplitude $\Gamma' = \Gamma/W'$ (Fig. \ref{fig:Lorentzian_sum})
\begin{equation}
    \Gamma \frac{\Omega^2}{\left( \omega - \omega_k^\text{0} \right)^2 + \Omega^2} =
    \sum_{k=1}^{W'} \Gamma' \frac{\Omega^2}{\left( \omega - \omega_k^\text{0} \right)^2 + \Omega^2}.
\end{equation}
By doing so, we are diminishing the $\Gamma'/\Omega$ ratio and giving more time to the pseudomode before active-space saturation effects are visible.

In Fig. \ref{fig:more_better} in the main text, we show how the number of pseudomodes influences the dynamics of the site-1 population of an excitonic dimer.
The results have been obtained by performing classical simulations of eq. \ref{eq:master_equation} using \texttt{qsextra} module which exploits \texttt{qutip} master equation propagator \cite{Johansson2013} (see ``Example 3" in Ref. \citenum{QSExTra}).
For the dimer, we set $\epsilon_1 = \epsilon_2 = 0$, $J_{12} = 1$ and initial population localized on site 1.
The environment of each chromophore is described by a Lorentzian spectral function with $\omega_k^\text{0} = 0$, $\Gamma = 20 J_{12}$ and $\Omega = 0.1 J_{12}$.
We show simulations of $W'=1$ pseudomode with $d=16$ states (blue line), $W'=2$ pseudomodes with $d=4$ states each (orange line), and $W'=4$ pseudomodes with $d=2$ states each (green line).
Notice that, in all the cases, the environment of one chromophore requires 4 qubits to be represented in the quantum register.
As a reference, we propagate the system using \texttt{qutip} implementation of the 
Hierarchical Equations Of Motion (HEOM) (red line).
The comparison with the reference shows a perfect agreement with the simulation that uses 4 pseudomodes.

In conclusion, we argued that under certain circumstances where the environment is highly non-Markovian, better performance can be achieved by using $W'$ pseudomodes with $d=2$ states, rather than using a single pseudomode with $2^{W'}$ states.
While this heuristic approach is not always valid, we observed diverse realizations where it holds.
The rationale will be the focus of further investigations.
%
%%%%%%%%%%%%%%%%%%%%%%%%%%%%%%%%%%%%%%%%%%%%%%%%%%%%%%%%%%%%%%%%%%%%%%%%
%%%%%%%%%%%%%%%%%%%%%%%%%%%%%%%%%%%%%%%%%%%%%%%%%%%%%%%%%%%%%%%%%%%%%%%%
\section{\label{app:thetas}Rotation angles in Figure \ref{fig:circuit}}
The gates appearing in Fig. \ref{fig:circuit} of the main text are defined as
\begin{equation*}
\begin{split}
    &R_z\left(\theta\right) = \exp{-i\frac{\theta}{2}\sigma^z},\\
    &R_x\left(\theta\right) = \exp{-i\frac{\theta}{2}\sigma^x},\\
    &R_{xx}\left(\theta\right) = \exp{-i\frac{\theta}{2}\sigma^x \otimes \sigma^x},\\
    &R_{yy}\left(\theta\right) = \exp{-i\frac{\theta}{2}\sigma^y \otimes \sigma^y},\\
    &R_{zx}\left(\theta\right) = \exp{-i\frac{\theta}{2}\sigma^x \otimes \sigma^z},
\end{split}
\end{equation*}
with rotation angles
\begin{equation*}
\begin{split}
    &\theta_1 = - \epsilon_1 \frac{\Delta t}{n_{\text{T}_1}},\\
    &\theta_2 = - \epsilon_2 \frac{\Delta t}{n_{\text{T}_1}},\\
    &\theta_3 = - \omega^0 \frac{\Delta t}{n_{\text{T}_1}},\\
    &\theta_4 = J_{12} \frac{\Delta t}{n_{\text{T}_1}},\\
    &\theta_5 = g \frac{\Delta t}{n_{\text{T}_1}},\\
    &\theta_6 = - g \frac{\Delta t}{n_{\text{T}_1}},\\
    &\theta_7 = \sqrt{2 \Omega \Delta t}  \frac{1}{n_{\text{T}_2}},\\
\end{split}
\end{equation*}
where we recall that $g = \sqrt{\Gamma \Omega/2}$.

Note that, if the pseudomodes are two-level systems ($d=2$), as in the example, one can always set $n_{\text{T}_2} = 1$ as gates $R_{xx}$ and $R_{yy}$ do always commute.
\bibliography{bibliography}

\end{document}